\RequirePackage{fix-cm}
\documentclass[smallextended]{svjour3}       
\smartqed  
\usepackage[utf8]{inputenc}
\usepackage{graphicx}
\usepackage{multirow}
\usepackage{natbib} 
\usepackage{pdflscape}
\usepackage{afterpage}
\usepackage{capt-of}
\usepackage{longtable}
\usepackage{lscape}
\usepackage[pass]{geometry}
\usepackage{tikz}
\usepackage{hyperref}
\graphicspath{ {img/} }
\begin{document}

\title{Student Success Prediction in MOOCs
}

\author{Josh Gardner         \and
        Christopher Brooks 
}


\institute{Josh Gardner \at
              School of Information \\
              University of Michigan\\
              \email{jpgard@umich.edu}           
           \and
           Christopher Brooks \at
              School of Information \\
              University of Michigan\\
              \email{brooksch@umich.edu}   
}

\date{Received: date / Accepted: date}

\maketitle


\begin{abstract}
Predictive models of student success in Massive Open Online Courses (MOOCs) are a critical component of effective content personalization and adaptive interventions. In this article we review the state of the art in predictive models of student success in MOOCs and present a categorization of MOOC research according to the predictors (features), prediction (outcomes), and underlying theoretical model. We critically survey work across each category, providing data on the raw data source, feature engineering, statistical model, evaluation method, prediction architecture, and other aspects of these experiments. Such a review is particularly useful given the rapid expansion of predictive modeling research in MOOCs since the emergence of major MOOC platforms in 2012. This survey reveals several key methodological gaps, which include extensive filtering of experimental subpopulations, ineffective student model evaluation, and the use of experimental data which would be unavailable for real-world student success prediction and intervention, which is the ultimate goal of such models. Finally, we highlight opportunities for future research, which include temporal modeling, research bridging predictive and explanatory student models, work which contributes to learning theory, and evaluating long-term learner success in MOOCs. 

\end{abstract}

\section{Introduction}

In their short history to date, Massive Open Online Courses (MOOCs) have simultaneously generated enthusiasm, participation, and controversy from both traditional and novel participants across the educational landscape. Trying to understand and improve enrollment, completion, and the overall learner experience has led to efforts to generate effective student models which can predict student dropout, completion, and learning in MOOCs. Despite the extensive attention devoted to such work by several related research communities and by the popular media, little overall synthesis of this work has been performed. We believe that such a synthesis is necessary, now more than ever, for several reasons.

\begin{figure}
    \centering
    \includegraphics[width=\textwidth]{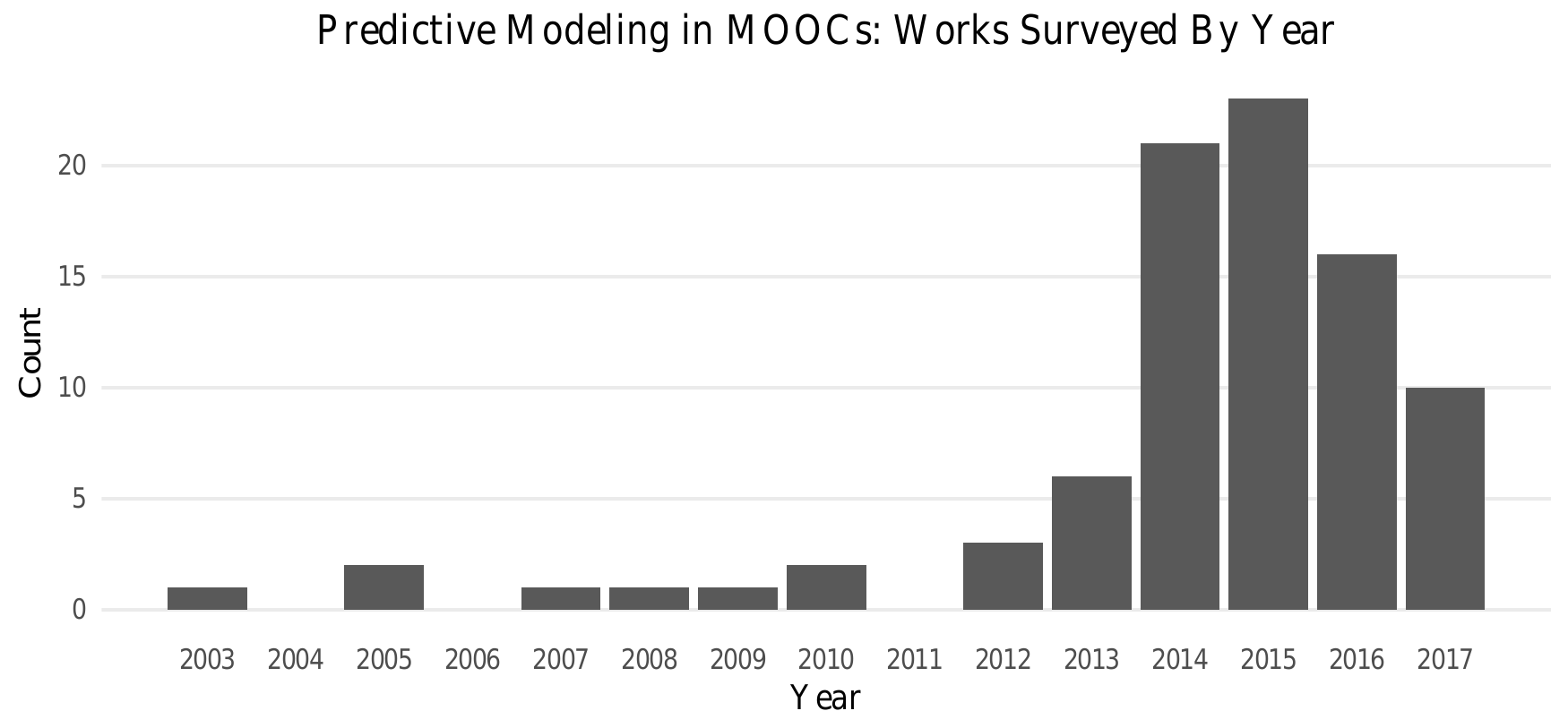}
    \caption{Published predictive modeling research in MOOCs over time. MOOC research has expanded dramatically since 2012, but little overall synthesis of predictive modeling work has been published. Even less work has synthesized or critically evaluated the feature extraction, modeling, and methodology of prior research, as we do in the current work.}
    \label{fig:years}
\end{figure}

First, MOOC research is at a critical stage in its development. An abundance of research has explored the phenomenon of MOOC dropout from several perspectives since the ``year of the MOOC'' in 2012 \citep{Pappano2012-bg}, as shown in Figure \ref{fig:years}. We survey $n = 87$ such studies in this work. A clear synthesis of this research is necessary in order to explore where consensus has emerged across the research community, where there may be research gaps or unanswered questions, and what action needs to be taken as a result of both. If we fail to learn from the lessons of several years of MOOC analysis, MOOCs may fail to deliver on their promise for millions of learners around the globe.

Second, there is a need to evaluate not only the findings of such research, but also its \textit{methodology}. Now that a body of research on student success prediction in MOOCs has accumulated, it is possible and appropriate to survey the techniques most commonly used. Such a critical survey allows us to disseminate consensus findings on effective techniques for student success prediction, to understand whether gaps exist, and to determine a future research agenda to address them. In particular, this issue is relevant to predictive modeling of student success in MOOCs because of the diverse communities that its practitioners are drawn from: education and the learning sciences, computer science, statistics and machine learning, behavioral science, and psychology researchers each bring different methods to the field. A methodological survey allows scientists to ensure that their knowledge is constructed on a strong methodological foundation, and to strengthen it where appropriate. In particular, this survey of predictive modeling allows for (a) the sharing of feature extraction and modeling approaches known to be effective, while also encouraging exploration into under-researched methods, and (b) sharing of overarching experimental protocols, such as prediction architectures and statistical evaluation techniques, which affect the inferences such modeling experiments produce. In this work, we provide detailed and novel data about the state of predictive student modeling in MOOCs for researchers interested in both (a) and (b).

Third, a critical promise of student success prediction in MOOCs has not yet been delivered on: the use of these predictions to actively improve learner outcomes and experiences through the operationalization of predictive models in MOOC platforms. We hope that this work can identify effective strategies for such tools to be used ``in the wild'' in active courses to achieve their oft-stated goal of impacting learner success in MOOCs. The implementation of live, real-time tools and personalized interventions stands to benefit from effective predictive modeling which can target and personalize interventions for those who need them most. Additionally, the implementation of predictive modeling as part of a MOOC has never been more practically achievable, as both the hardware and software required for user modeling in digital environments (such as MOOCs) have become increasingly accessible. The use of predictive models for adaptive user experiences more broadly has grown quite common, and is commonly executed at a massive scale (for example, prediction-based targeted advertising on the World Wide Web). A clear knowledge of the research consensus on effective predictive modeling methods in MOOCs will support the construction of such tools, effectively ``closing the loop'' of predictive modeling in MOOCs. We leave the development of the interventions based on these predictive models to future work.

In the work that follows, we address each of these three goals. In the remainder of this section, we provide the reader with a basic introduction to MOOCs and survey the state of the overall MOOC landscape to date. In Section \ref{sec:mooc-prediction}, we dive deeper into the specific focus of this work by discussing student success prediction in MOOCs, introducing the task, the data typically available for its execution, and the basic procedure for the construction and evaluation of predictive models. Section \ref{sec:lit-review} surveys prior work on predictive models of student success in MOOCs, including a detailed matrix of $n = 87$ previous works on this topic in Table \ref{tab:master-matrix}. We synthesize the results of this survey in Section \ref{sec:synthesis}, highlighting overall trends and providing detailed data on the methodologies used across the sample of works surveyed. We discuss research gaps, methodological issues, and unanswered questions suggested by the literature survey in Section \ref{sec:meth-res-gaps}. Opportunities for future research suggested by our survey, as well as our interpretation of the direction of the field, are discussed in section \ref{sec:future-research}. We conclude in Section \ref{sec:conclusion}. 

This work is part of a series on predictive models in MOOCs, and in future works we provide a discussion of techniques for model evaluation, and infrastructure for replication of machine learned models in MOOCs.

\subsection{MOOCs: A Novel Educational and Research Context}\label{sec:mooc-intro}

Massive Open Online Courses are enticing, in part, because they are so different from many other forms of education. However, exactly what a MOOC \textit{is} is itself the subject of some debate. We do not seek to fully resolve this debate here, but in this section, we detail several generally agreed-upon characteristics of MOOCs in order to build a working definition for use within this review. 

We take MOOCs to have the following attributes:

\textbf{Massive, Open and Online:} By definition, these are the attributes most closely associated with MOOCs. MOOCs are \textit{massive} in that they typically have far more students than even the largest traditional classroom courses. This would include, at minimum, hundreds of learners for specialized courses to hundreds of thousands of learners for more general or popular courses. The instructional team tasked with supporting these learners is typically very small; therefore the student-teacher ratio in these courses is far higher than in traditional higher education or e-learning courses. MOOCs are \textit{open} to all learners, often being both public and free. The two largest English-language MOOC providers, Coursera and edX, initially offered all courses free of cost, though business model changes have seen more barriers to taking free courses over time (though both platforms still offer financial aid programs, and at least partial access for unpaid learners in most courses). The openness of MOOCs is perhaps what makes them most exciting by providing access to high-quality educational experiences for all learners around the globe.\footnote{Other work has emphasized the ``openness'' of MOOCs as reflective of open \textit{content} and open-ended \textit{learning structures} \citep[e.g.][]{Kennedy2015-jw}; this is highly debatable with current MOOC providers, where much of the content is under copyright and may follow strict instructivist designs, and we consider these senses of openness to be too constraining for the present work.} Finally, MOOCs are \textit{online} -- they are digital, internet-based courses, not in-person courses. Course materials, assignments, instructors, and peers are all accessed on the World Wide Web via a computer or other device with a web browser or a dedicated platform-specific application.

\textbf{Low- or No-Stakes:} Traditional higher education and e-learning courses are typically taken strictly for academic credit or other official certifications, often at a non-trivial financial cost to the participant, with implicit or explicit penalties for poor performance (e.g. low grades, loss of tuition without credit). In contrast, MOOCs provide the option to simply take the course independent of any certification, credit, or degree program, with no penalty for repeating or failing to complete the course. This gives MOOCs a particularly unique set of course participants who sometimes have little or no investment in completing a course, making the task of student success prediction (and, consequently, the task of student support based on these predictions) particularly challenging. Under this definition, paid and for-credit online courses are typically not considered MOOCs. Many other low- or no-stakes learning environments exist -- such as textbooks and tutorials, museums, and other offline and online resources -- but these environments do not share the other features of MOOCs.

\textbf{Asynchronous:} The time scale for content consumption and participation in a MOOC tends towards the flexible, although the degree of this flexibility may vary. Many MOOCs are clearly divided into ``modules,'' often by week, which are released to learners over time. These courses often have clearly-defined start and end dates, with successful completion being contingent upon learners meeting specified criteria by the course end date. Within these time windows, however, learners were typically free to browse content and complete assignments in any order and at any time. A fully asynchronous model has recently become more common in MOOCs, where learners have access to all content on demand after entering a course, and can complete content at their own pace. We note that this model has coincided with the transition to a subscription-based, as opposed to course-based, pricing model on certain MOOC platforms.

\textbf{Heterogeneous}: As a direct consequence of many of these features, the population of learners in MOOCs is heterogeneous in terms of both demographics and intentions \citep{Koller2013-zk, Chuang2016-oj}. Even as course populations skew toward college-educated males from industrialized countries, these course populations are still far more diverse than any of the other educational contexts superficially similar to MOOCs \citep{Glass2016-sz}. The backgrounds of learners vary significantly, from graduate-level educated learners who are employed full-time in the subject area of the course, to students without a high school diploma. Learners vary in gender, age, nationality, and intent. The majority of MOOC students are located outside the United States and hold a bachelor's degree \citep{Chuang2016-oj}, and there is also evidence that teachers are well-represented in course populations \citep{Seaton2015-as, Chuang2016-oj}. However, obtaining even basic demographic data on users is currently only available through on optional questionnaires with low response rates \citep{Kizilcec2015-gu, Whitehill2015-ap, DeBoer2013-ki}. As a result, predictive models are often unable to utilize this data directly and instead need to draw directly on learner behavior, not demographics or reported intentions.

Together, these features of MOOCs define an educational environment that is sufficiently different from other well-studied environments -- such as e-learning, on-campus higher education, or digital K-12 education -- to justify the formation of a new and separate predictive modeling literature. As an illustrative example, consider a comparison of a ``dropout'' student (a non-completer) in a MOOC versus any of the traditional contexts mentioned above. One might reasonably expect different factors to contribute to dropout, different subpopulations to be most likely to drop out, and for  learners to experience different consequences of dropping out, in a MOOC compared to other educational contexts. Indeed, \cite{DeBoer2014-xl} argues for a broad reconceptualization of traditional student success metrics in MOOCs instead of the use of terms grounded in traditional education courses, such as the term ``dropout;'' \cite{Reich2014-ob} proposes ``stopout'' as a more appropriate term for this outcome. 

As we will discuss below, there are also very different data sources available in MOOCs compared to other educational contexts: for example, MOOCs collect rich, granular behavioral data at a level that is unavailable in almost any other context. MOOCs are also characterized by a lack of complete and reliable historical or demographic data; in contrast, institutional course providers (such as brick-and-mortar schools) typically lack any readily-available behavioral data but have rich historical, demographic, and co-curricular data. These data sources are directly relevant to the predictive models which they are used to construct in each context. Again, this implies a material difference between predictive modeling in MOOCs and other educational environments.

Our goal in describing these features of MOOCs is not to argue for a particular conceptualization of MOOCs; it is simply intended to introduce the basic concept of a MOOC to readers, and to motivate the features of MOOCs used as the criteria for inclusion in the literature review in Section \ref{sec:lit-review} below. 

\subsection{The State of the MOOC Landscape}~\label{sec:mooc-state}

As of 2017, an estimated 81 million students have registered for or participated in at least one MOOC \citep{Shah2018-cn}. The five largest MOOC providers, according to self-reported enrollment numbers, are Coursera\footnote{\url{https://www.coursera.org/}}, 30 million registered users; edX\footnote{\url{https://www.edx.org/}}, 14 million registered users; XuetangX\footnote{\url{http://www.xuetangx.com/}},  9.3 million registered users; Udacity\footnote{\url{https://www.udacity.com/}}, 8 million registered users; and FutureLearn\footnote{\url{https://www.futurelearn.com/}}, 7.1 million registered users \citep{Shah2018-cn}. Enrollment continues to grow over time, but there is some indication that enrollments have begun to slow as platforms have transitioned to paid models and phased out various free certification options, and as the course population declines in size over repeated iterations of a course \citep{Chuang2016-oj}.

These impressive enrollment figures mask a well-known issue with the MOOC experience: around 90\% of students who enroll in a MOOC fail to complete it \citep{Jordan2014-go}. Given the lack of barriers to entry, massive course populations, and high student to teacher ratios in MOOCs, this may not be particularly surprising. As shown in Table \ref{table:outcome-metrics}, a majority of predictive modeling research in MOOCs has focused on dropout prediction. While the massive dropout rate may fail to account for student intentions \citep{Koller2013-zk}, the best data indicates that slightly more than half of students intend to achieve a certificate of completion in a typical MOOC, and around 30\% of these respondents achieve this certification \citep{Chuang2016-oj}. This low completion rate even among intended completers is still cause for concern. Effective predictive models can support several approaches to improving MOOC dropout rates.

As of 2017, MOOCs cover a variety of topics, with over 6,850 courses offered by more than 700 universities across these platforms \citep{Shah2018-cn}. Coursera, for instance, offers more than 180 specializations (sequences of courses in a specific topic area, such as ``Data Structures and Algorithms'' or ``Dynamic Public Speaking''). There are several full online graduate degrees offered on the platform, such as the Master of Business Administration iMBA program offered by the University of Illinois, Urbana-Champaign on Coursera. The edX platform offers pathways for learners into higher education, such that when a program (called a MicroMasters) is completed on the MOOC platform, learners are then provided with credit transfer if they subsequently enroll in a residential graduate program. The University of Arizona's Global Freshman Academy provides the opportunity for students to complete their entire freshman year online. Regardless of platform, format, and structure, Computer Science courses continue to be the most popular courses on the platform, with science, history, business, and health courses also popular \citep{Chuang2016-oj, Shah2018-cn, Whitehill2017-tt}.

\section{Student Success Prediction in MOOCs}\label{sec:mooc-prediction}

Before surveying the vast body of prior work on student success prediction in MOOCs, in this section we seek to clearly define and motivate the task. This framing is essential to the discussion below and to the conclusions we draw from this review.

\subsection{Defining Student Success}

Student success in a MOOC can be viewed from several different perspectives. Several outcomes have been used to measure and predict student success in MOOCs, including completion, certification, overall course grades, and exam grades, shown in Table \ref{table:outcome-metrics}. The task of discussing student success in MOOCs is particularly challenging due to the fact that we typically apply language and metrics adopted from traditional educational settings -- i.e., dropout, achievement, participation, enrollment -- that can mean different things, or seem incoherent, in the context of a MOOC \citep{DeBoer2014-xl}. 

In the context of this work, we define student success as encompassing a broad class of metrics which measure course completion, engagement, learning, or future achievement related to the content or goals of a MOOC. We believe that each of these broad categories suggests at least one kind of motivation participants in a MOOC might have for joining the course, but each alone is certainly inadequate to describe ``success.'' We review work which presents the results of a predictive model of any type of student success according to this definition.

Having several potential metrics to describe student success in MOOCs is useful for several reasons:  (a) it allows us to capture metrics related to the diverse goals MOOC learners have, such as course completion, certification, career advancement, or subject mastery \citep{Koller2013-zk, Reich2014-ob}; (b) it reflects the lack of research consensus on how to measure student success in MOOCs \citep{Perna2014-cp, DeBoer2014-xl}; and (c) it allows us to test the robustness of models by potentially checking their ability to predict multiple different outcomes. While (c) has been an uncommon approach to date, we believe that this is an important avenue for future work (for one example, see \cite{Fei2015-ea}).

Several metrics are used to measure student success in the works surveyed below. A collection of the most common metrics used for student success prediction and the frequency with which they occur in our literature review is shown in Table \ref{table:outcome-metrics}. For an examination of alternative long-term metrics of student success, see \cite{Wang2017-br}.

\begin{table}[!t]
\centering
\begin{tabular}{ l p{6cm} p{1cm}}
\hline
\textbf{Outcome} & \textbf{Description} & \textbf{Count} \\ \hline
\textbf{Dropout} & A student drops out if they do not ``complete'' a course. Often operationalized as whether a student continues participating in a course until the course concludes. & \multirow{2}{*}{39} \\
\textbf{Stopout} & A student is a ``stopout'' if they stop engaging with the course prior to the end of the course \cite{Taylor2014-hu}. Often, stopout is functionally equivalent to dropout, but merely emphasizes that we often cannot observe a students' intention to ``drop out'' and instead only observe whether they stop interacting with the course. & \\
\textbf{Certification} & Certification is achieved when a student earns a certificate of accomplishment for the course. This typically consists of earning enough points on course assignments to meet some pre-determined threshold for the course (typically around 70\%). & 16 \\
\textbf{Final Exam Grade} & The students' grade on the course final exam, a cumulative test typically administered near the end of the course which tests the full subject matter taught in the course. & 4 \\
\textbf{Final Course Grade} & The students' overall grade in the course, calculated based on the instructors' specification. Typically, course grades are a mix of one or more of the following: in-video quizzes, out-of-video quizzes, homework assignments, problem sets, human-graded assignments, and exams. & 14 \\
\textbf{Pass/Fail} & A student typically passes a course if they meet or exceed an instructor-specified overall grade threshold; otherwise they fail. & 7 \\ 
\textbf{Other} & Correct on First Attempt (CFA) \cite{Brinton2015-ya}, increase/decrease in engagement \cite{Bote-Lorenzo2017-yh}, etc. & 15 \\ 
\hline
\end{tabular}
\caption{Common student success metrics for predictive models in MOOCs. Dropout/stopout prediction is the most common, but several learning outcomes (final exam grade, course grade, pass/fail) have also generated significant attention.}
\label{table:outcome-metrics}
\end{table}

\subsection{Why Model Student Success in MOOCs?}

Student success predictions are useful for a wide variety of tasks, and these models vary along three main dimensions relevant to these tasks (shown in Figure \ref{fig:modeling-dimensions}). We identify three main reasons for developing predictive models of student success:

\begin{figure}[]
    \centering
    \includegraphics[width = \textwidth]{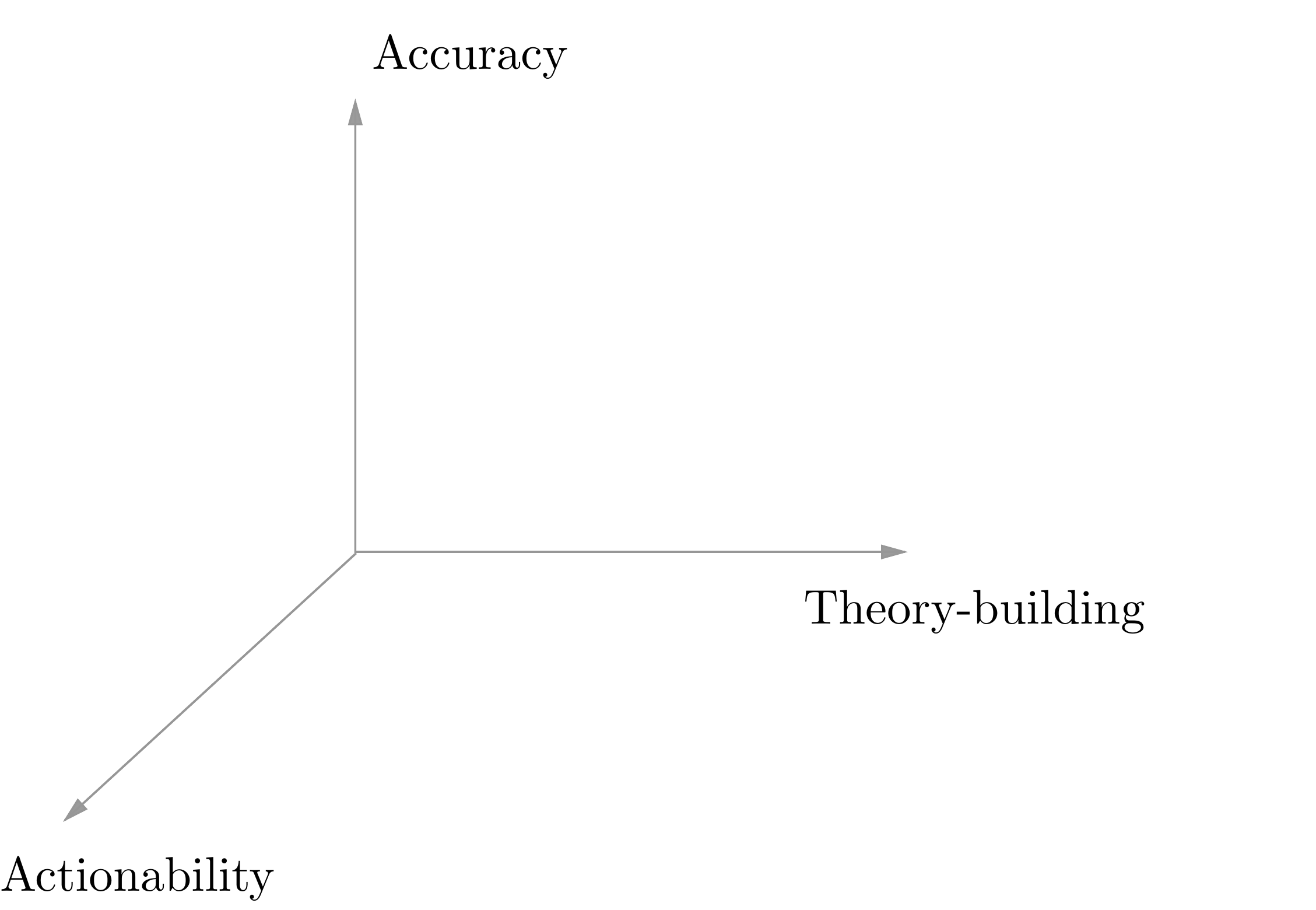}
    \caption{Three salient dimensions of predictive models in MOOCs. Models vary along all three dimensions, but there is no strict trade-off between any dimensions. We synthesize the state of MOOC research with respect to these dimensions, and highlight methodological gaps needed to improve predictive student models, in Section \ref{sec:meth-res-gaps}.}
    \label{fig:modeling-dimensions}
\end{figure}

\textbf{Personalized Support and Interventions:} Identifying students likely to succeed (or not succeed) has the potential to improve the student experience by providing targeted and personalized interventions to those students predicted to need assistance. This is the stated motivation behind much of the work surveyed here, which often refer to these students as ``at risk'' learners (a term adopted from the broader educational literature). In particular, because of the massive student population in MOOCs relative to the size of the instructional support staff, clearly identifying struggling students is important to providing those students with targeted and timely support. Many of the ``human'' resources in MOOCs are quite scarce (i.e., instructor time), and predictive models can provide timely guidance on (a) \textit{identification} of which students need these resources, and (b) \textit{intervention} by predicting which resources can best support each at-risk student. While a teacher might be able to directly observe students in a traditional in-person higher education course, or even in a modestly sized e-learning course, such observation is not available to support MOOC instructors at scale, and predictive models can serve this purpose. Particularly when instructor time and resources are scarce, predictive models which can identify these students with high confidence and accuracy are required. Additionally, many interventions would be unnecessary or even detrimental to the learning of engaged or otherwise successful students.

In order to deliver personalized support and interventions, a predictive model must provide predictions which are both \textit{accurate} and \textit{actionable}. We refer to the dimension along which model predictive performance varies in its ability to relate student behavior or attributes to the outcome of interest as its \textit{accuracy}. We discuss how to measure the quality of a model's predictions in Section \ref{sec:mod-eval}. Here, it suffices to say that accuracy is critical to the delivery of personalized interventions; a model which cannot correctly identify students at risk of dropout cannot effectively support interventions to prevent it. Furthermore, the predictions of such a model must also be \textit{actionable}. That is, these predictions must enable targeted and timely interventions for supporting student success. We argue in Section \ref{sec:meth-res-gaps} that there are problems with the actionability of most prior predictive modeling research in MOOCs due to their prediction architecture, which often cannot be implemented in actively running courses.

\textbf{Adaptive Content and Learner Pathways:} Predictive models in MOOCs stand to enable the delivery of course content and experiences in a way that optimizes for expected student success. Very little prior research has utilized adaptivity or true real-time intervention based on student success predictions of any form in MOOCs. \cite{Whitehill2015-ap} utilizes dropout prediction to optimize learner response to a post-course survey (this work optimizes for data collection, not learner success), and \cite{He2015-ab} describes a hypothetical intervention based on predicted dropout probabilities (but only implements the predictive model to support it, not the intervention itself). \cite{Kotsiantis2003-km} describes a predictive model-based support tool for a distance learning degree program of 354 students, a scale far smaller than most MOOCs. The work which most clearly demonstrates adaptive content and learner pathways of which the authors are aware is \cite{Pardos2017-lh}, which implements a real-time adaptive content model in an edX MOOC. However, this implementation is optimized for time-on-page, not student learning. The dearth of research on adaptive content and learner pathways supported by accurate, actionable models at scale is, at least in part, due to a lack of consensus on the most effective techniques for building predictive models in MOOCs, which we address through the current work.

\textbf{Data Understanding:} Predictive models can also be useful \textit{exploratory} or \textit{explanatory} tools that help understand the mechanisms behind the outcome of interest. Instead of strictly providing predictions to enable personalized interventions or adaptive content, predictive models can be tools to identify learner behaviors, learner attributes, and course attributes associated with success in MOOCs. These insights can drive improvements to the content, pedagogy, and platform, and contribute to our understanding of the underlying factors influencing student success in these contexts. They also contribute more directly to theory by providing a more detailed understanding of the complex relationships between predictors and outcomes discovered via predictive modeling. We describe this dimension of models as \textit{theory-building} to highlight their usefulness in the formation of theories about these underlying factors. From this perspective, certain types of models are more useful than others: models with straightforward, interpretable parameters (such as linear or generalized linear models, which provide interpretable coefficients and $p$-values; and decision trees, which generate human-readable decision rules) are far more useful for human understanding of the underlying relationship than those with many complex and interacting parameters (such as a multilayer neural network). Unfortunately, the latter are usually (although not always) more effective in making predictions in practice, so there is often a tradeoff between interpretability and predictive performance. Recent advances in making more complex models interpretable suggest that this tradeoff may be reduced in the future \citep[e.g.][]{Baehrens2010-pa, Craven1996-gq, Ribeiro2016-ut}, but at present this ``fidelity-interpretability tradeoff'' is still a salient issue for predictive models in MOOCs \citep{Nagrecha2017-dn}. This issue is further discussed in Section \ref{sec:two-cultures} below.

\subsection{Data for Student Success Prediction in MOOCs}~\label{sec:mooc-data}

In this subsection, we briefly describe the raw data available for student success prediction in MOOCs, including the common formats, schema, and types of behaviors and metrics collected. We provide data on the use of each raw data source across works surveyed in Section \ref{sec:synthesis}.

Student success prediction in MOOCs has attracted a great deal of enthusiasm in part because of the data available to researchers interested in studying MOOCs. Digital learning environments such as MOOCs provide rich, high-granularity data at a scale simply not available in traditional educational contexts. While this data varies slightly from platform to platform, because of the dominance of only a few large MOOC providers (most notably, Coursera and edX), the available datasets are remarkably consistent in practice. This is useful for several reasons: (a) enables the use of consistent feature extraction and modeling methods, even across platforms, which reduces both development and computation time; (b) it allows for direct replication of research across courses and even across platforms \citep{Gardner2018-hj}.

Common data generated by MOOC platforms are discussed below. The frequency with which these data types were utilized across our literature survey is shown in Figure \ref{fig:data_source} below.
\\

\subsubsection{Clickstream Exports}

\begin{figure}
    \centering
    \includegraphics[width=\textwidth]{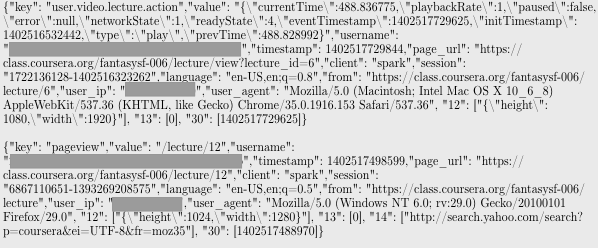}
    \caption{Sample clickstream entries, sensitive data redacted for publication.}
    \label{fig:clickstream-sample}
\end{figure}

Clickstream exports, also called server logs or clickstream logs, are typically records of every interaction with the server which hosts the course platform in JavaScript Object Notation (JSON) format. These interactions include every request to the web server hosting the course content, including each mouse click, page view, video play/pause/skip, question submission, forum post, etc. The same metadata is recorded for each interaction, and from this record, we can build detailed datasets at several levels of aggregation. An example of entries from a clickstream log is shown in Figure \ref{fig:clickstream-sample}; note the many detailed attributes recorded for each interaction. Clickstream exports are the most raw, high-granularity data available from MOOC platforms. However, this granularity also presents a challenge: raw clickstream data cannot be directly used as input for most predictive models; instead, ``features'' -- attributes relevant to the outcome of interest -- need to be manually \textit{extracted} from the clickstream log. This is a labor-intensive process (we use the terms \textit{feature engineering} and \textit{feature extraction} interchangeably to refer to this process). Feature engineering appears more important to the effectiveness of predictive models than the statistical algorithm itself (see Section \ref{sec:lit-review} for a more detailed discussion of the importance of feature engineering). Indeed, many of the works surveyed here introduce innovations only to the feature engineering method and adopt otherwise standard classification algorithms for predicting student success from clickstream data \citep[e.g.][]{Brooks2015-ej, Veeramachaneni2014-ug}).

Clickstream data also presents a challenge of scale. This data is often quite large (tens of gigabytes for a single course), due to its granular nature and the many individual interactions that take place over the duration of a MOOC. Any aggregation of individual user sessions or interactions requires manually parsing and aggregating data from the clickstream. Simply reading, processing, and extracting the features from such data can be computationally expensive. \\

\subsubsection{Forum Posts}

\begin{figure}
    \centering
    \includegraphics[width=5cm]{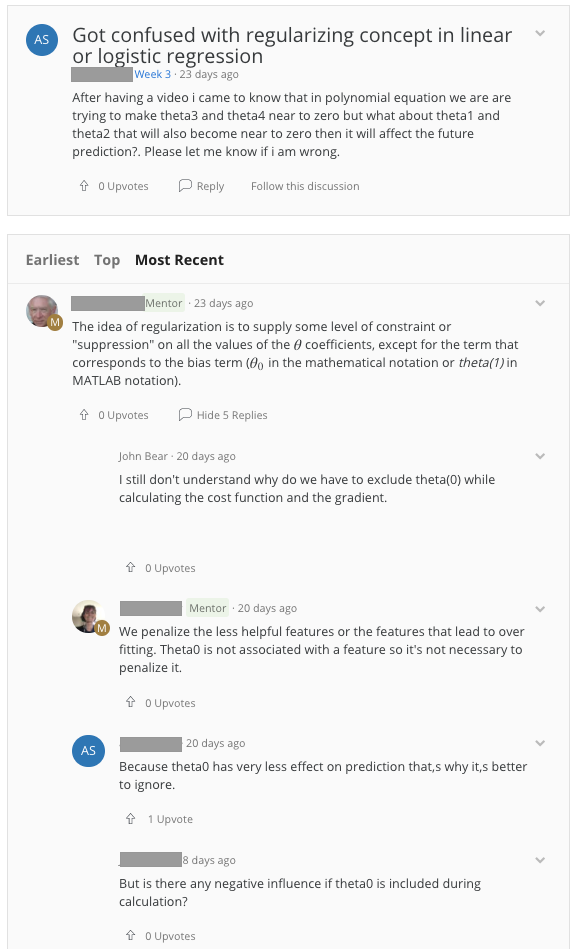}
    \caption{An example of a threaded forum post in a Coursera MOOC. Visible are the user-generated text, threaded replies (note that some are hidden from this view), and optional upvotes.}
    \label{fig:forum_thread}
\end{figure}

A defining feature of most MOOCs is a set of thread-based discussion fora used for various tasks, including interactions directly related to course content and more general community-building and discussion. Different platforms implement discussion fora differently\footnote{DiscourseDB (\url{http://discoursedb.github.io/}), and MOOCdb \footnote{\url{https://github.com/MOOCdb}} are both tools used to bridge these different implementations and data sources across platforms to enable research and encapsulate the full breadth of forum experiences. Both are now components of LearnSphere (\url{http://learnsphere.org/})}, but across every major platform, the text of forum posts and a variety of metadata and related interactions (such as upvotes for questions or answers) are typically collected in a relational database, accessed via Structured Query Language (SQL).  As shown in Figure \ref{fig:data_source}, forum post data is second only to clickstream data in terms of its use in predictive models of student success in MOOCs. This data is often used to extract (a) measures of engagement, by tracking users' forum viewing patterns; (b) measures of mastery, understanding, or affect, generated by applying natural language processing to the raw text of forum posts; and (c) social network data by assembling graphs where various connections in the fora constitute edges. An illustration of a threaded discussion post in a Coursera course is shown in Figure \ref{fig:forum_thread}. \\

\subsubsection{Assignments}

Assignments are often used in MOOCs similar to the way they are used in residential or in-person courses, and data related to assignment submission is also often stored in a relational database. A variety of assignment types are used in MOOCs, including automatically graded assignments (such as multiple-choice assessments and small programming tasks), manually-graded assignments (such as data analysis reports or essays, which can be graded by both course instructors or, more commonly, peers in the course), in-video questions, interactive lab simulations, and programming assignments completed in external environments (e.g., Jupyter notebooks). Assignment data is typically limited to metadata (i.e., open date, due date) and assignment-level or (less commonly) question-level data about submissions or data about the content of submissions (such as text cohesion metrics of written work or syntactic analysis of submitted code). As Figure \ref{fig:data_source} indicates, the use of assignment features is less common, likely due to a combination of (a) the low number of users who complete assignments in MOOCs, as a proportion of total registrants or participants, and (b) the substantial variation across courses in the way assignments are used. \\

\subsubsection{Course Metadata}

Detailed information about the course and instructional materials are also typically recorded in MOOC platforms and retained for \textit{post-hoc} analysis. This includes information about course modules, video lectures (length, title, module), and assignments (including quizzes, homework, essays, human-graded assignments, exams, etc.). Little research has actively explored the use of course metadata in predicting student success. The research which has evaluated such data, however, suggests that it may indeed impact factors such as learner persistence and engagement \citep[e.g.][]{Evans2016-gj, Qiu2016-ct}. \\

\subsubsection{Learner Demographics}

Most MOOC platforms also record information about learner demographics, when it is available. However, such information is typically collected via optional pre- and post-course surveys, which are subject to various response biases \citep{Kizilcec2015-gu}. While this information is potentially interesting, its limited availability (and bias in the data that is available) has limited the research on demographics in MOOCs to date to a small number of studies which we survey in Section \ref{sec:demographic-models}. \cite{Hansen2015-bc} explores using external datasets and IP address-based geolocation to fetch additional demographic data, but not for predictive student modeling.

\subsection{Relation to Other MOOC Research}

The predictive modeling research evaluated in this work is situated in the context of a much larger and broader body of MOOC-related research. Prior research on MOOCs has covered a broad variety of topics, including changes in learner discourse over time \citep{Dowell2017-ij}, interventions to improve student completion \citep{Kizilcec2017-sl}, demographics and participation rates and the relationship to course activity \citep{Guo2014-eh}, and student plagiarism and academic honesty issues \citep{Alexandron2017-kk}. Additionally, the researchers addressing this topic, both in the predictive context and more broadly, come from a wide variety of academic perspectives, including learning theory, social and experimental psychology, computer science, statistics, economics, design, and linguistics. 

Predictive modeling most often occurs in research contexts where the goal is either (a) data understanding (e.g., for learning theorists and psychologists with the aim of understanding the factors most closely associated with dropout) or (b) utilizing predictions as part of a larger learner support system which can be used to improve student experiences or outcomes (e.g., for instructional designers and platform architects). This distinction reflects a larger distinction between the ``two cultures'' of statistical modeling discussed in Section \ref{sec:two-cultures}. We consider both types of work (those focused on modeling for understanding, and those modeling for prediction) in this survey, as both contribute to the goals of understanding and supporting MOOC learners.

\section{Predictive Models of Student Success in MOOCs: A Feature, Outcome, and Model-Based Taxonomy}\label{sec:lit-review}

In this section, we survey prior research on predictive models of student success. We begin the review with an overview of our methodology and relevant categorizations, as well as our methodology and its motivation.

\subsection{Categorization Scheme}

This section describes the categorization scheme used to organize the literature review presented in this work. The three components used in the categorization are also defined in Table \ref{tab:categorization-scheme}.

\begin{table}[]
\centering

\begin{tabular}{l p{5.5cm} p{2.5cm}} 
\hline 
\textbf{Category} & \textbf{Definition} & \textbf{Example} \\ \hline 
\textbf{Features (predictors)}  & Structured data, typically extracted from raw MOOC platform data or collected using other means, which is used as the basis for a predictive model. & Count of forum posts; student gender. \\ 
\textbf{Outcome (prediction)}  & The label or outcome of interest of a predictive model on which model performance is evaluated. & Dropout status; final grade. \\ 
\textbf{Theory (model)}  & The conceptual or theoretical model which provides the basis for the hypothesis being tested by a predictive modeling experiment. & Social learning theory. \\  \hline 
\end{tabular}
\caption{Aspects of predictive modeling experiments used to categorize works surveyed.}~\label{tab:categorization-scheme}
\end{table}

\subsubsection{Feature-Outcome-Model Categorization}

The works below are grouped into broad conceptual categories based on the the input \textit{features}, the \textit{outcomes} of the prediction, and the theoretical \textit{models} used to motivate the work, when they are described. Generally, there is a strong association between these three components (i.e., experiments which use activity-based features most often predict an activity-based outcome, dropout, and are constructed to evaluate theories about learner behaviors; experiments using cognitive features most often predict a cognitive outcome, such as learning gains, and are supported by theories of cognition and learning). The strongest association is between the input features and the prediction outcome (as we will discuss in detail in Section \ref{sec:features-outcomes}). Theoretical motivations for predictive models are sometimes missing or left unstated (see Section \ref{sec:gaps-theory} for further discussion), but when these models are present, they often also align with the input data and the outcome of interest. While we note that the feature-model-outcome correlations are imperfect and there is significant overlap between many groups, we believe that this provides both an effective categorization of prior MOOC research as well as a reasonable model of how this research is conducted (with a set of input data, an outcome of interest, and a theoretical model or question about what is driving associations between input predictors and the outcome). Where a work fits into multiple categories, we discuss it in each applicable category below.

This categorization is a novel contribution of the current work, and has not been previously applied to predictive modeling research in MOOCs, to the authors' knowledge. Data describing the observed feature-outcome pairings  across prior research also contributes insight regarding well-researched areas, and gaps or opportunities for future research. For example, Table \ref{tab:features-outcomes} shows that only two works surveyed used performance-based features to predict course completion; further research in this area seems warranted.

Each of the broad model categories considered below has something important to offer predictive modeling efforts, but there are likely different underlying factors driving the predictive performance of student success in each category, which makes the separate discussion necessary. Similar feature-based groupings have been used or suggested in other works \citep[e.g.][]{Whitehill2015-ap, Whitehill2017-tt, Li2017-yd, Liang2016-us}.

\subsubsection{Feature Extraction as Critical to Predictive Modeling in MOOCs}

Feature extraction, in particular, emerged throughout our survey as a useful dimension on which to separate models, and an element of particular interest to predictive modeling researchers in MOOCs. It has been noted in several works that in addition to being perhaps the most difficult, feature extraction is also one of the most critical tasks in predictive models of student success \citep{Li2016-ze, Robinson2016-yr, Nagrecha2017-dn}. 

For example, \cite{Li2017-yd}, citing \cite{Zhou2015-zx}, notes that ``data preprocessing should be considered with more attention than learning algorithms''. \cite{Sharkey2014-tc} claims that feature extraction is ``arguably the most important step in the process of developing a predictive model.'' \cite{Taylor2014-hu} state that ``[w]e attribute success of our models to these variables (more than the models themselves)...any vague assumptions, quick and dirty data conditioning or preparation will create weak foundations for one’s modeling and analyses,'' emphasizing their feature extraction methods over their modeling techniques despite fitting over 70,000 models in this experiment. The same authors argue in \cite{Veeramachaneni2014-ug} that ``[h]uman intuition and insight defy complete automation and are integral part of the process'' of predictive modeling in MOOCs; they find that the most predictive features are complex, often relational (requiring the linking of multiple data fields), and were discovered through expert knowledge of both context and content. Feature extraction is highlighted as one of the core components of the dropout prediction problem in \cite{Nagrecha2017-dn}, which notes that ``the electronic nature of MOOC instruction makes capturing signals of student engagement extremely challenging, giving rise to proxy measures for various use-cases'' -- that is, the extraction of \textit{signal} (useful features) from the electronic records of a MOOC is a key task in the pipeline of predictive model-building. 

Therefore, we concluded that an effective categorization scheme for this review should highlight feature extraction techniques. The association between many feature extraction methods and the outcomes they are used to predict further ``brightens the lines'' of this categorization in many cases (such as with performance-based models, which are overwhelmingly used to predict academic performance as shown in Table \ref{tab:features-outcomes}). 

\subsubsection{Predictive Performance Evaluation}

Despite the current survey's emphasis on understanding predictive models of student success, we avoid categorizing the work surveyed based on their predictive results alone. This is because of large case-by-case variation in (a) the experimental subpopulations, which are different subgroups of different MOOC course populations, (b) the methodology and metrics for model evaluation, and (c) the outcome being predicted. These three factors are so divergent across the work surveyed that holding the performance of each experiment to the same standard would be more misleading than it would be useful, as we discuss below. 

Limited prior research has investigated the issue of how using different types of experimental protocols in predictive modeling experiments might influence or bias the results. This work has demonstrated how different prediction architectures, for example, can influence the results of predictive modeling experiments in MOOCs \citep{Boyer2015-lo, Brooks2015-ej, Whitehill2017-tt}.  We will discuss some of the methodological shortcomings that make conducting these comparisons so difficult in Section \ref{sec:meth-res-gaps} below, including inconsistent experimental populations; ineffective model evaluation; unrealistic or impractical prediction architectures; inconsistent model performance metrics; and others. In another work, we present a sociotechnical platform designed to enable direct replication of predictive modeling results on the same MOOC datasets, which can ameliorate the issue of ``apples-to-oranges'' model comparison faced by readers to date \citep{Gardner2018-hj}.

\begin{figure}
    \centering
    \includegraphics[width=\textwidth]{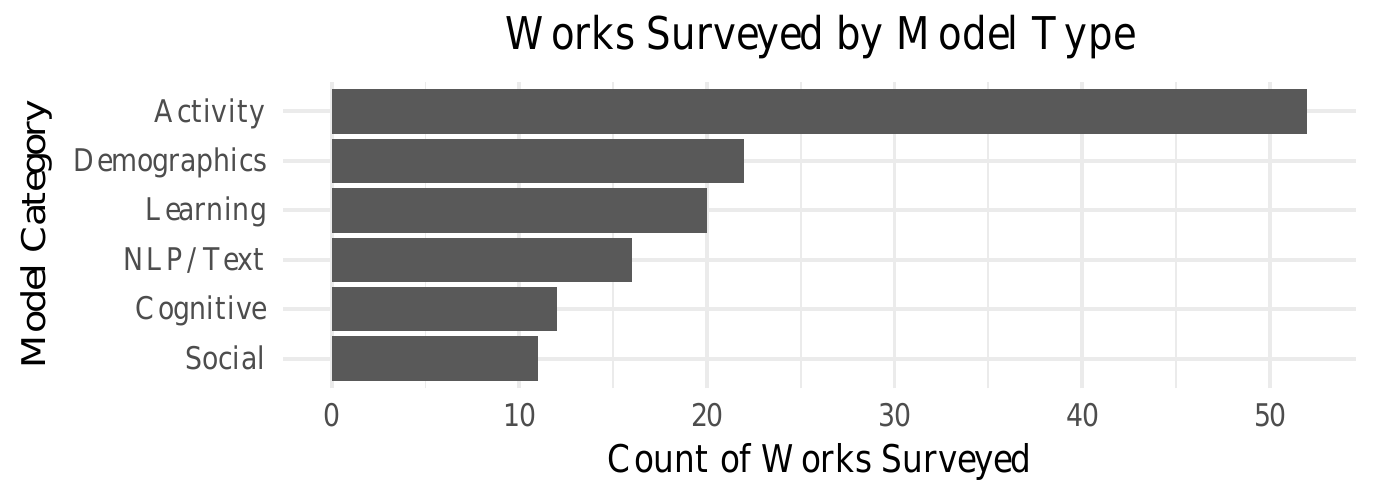}
    \caption{Counts of works surveyed by feature type, which broadly represents the most common approaches in predictive models of student success in MOOCs. Activity-based feature sets are the most common, which primarily reflects the activity-based outcome (dropout) most commonly predicted in the works surveyed. Note that experiments considering multiple types of features were counted in all relevant categories. Each category is defined in a corresponding subsection of Section \ref{sec:lit-review}.}
    \label{fig:model_categories}
\end{figure}

\subsection{Survey Methodology and Criteria for Inclusion}~\label{sec:survey-methodology}

We intend this to be a relatively broad, inclusive literature survey. We include work which (a) involves an application of predictive modeling of student success, where student success is broadly construed according to one or more of the metrics listed in Table \ref{table:outcome-metrics}; (b) doing so in the context of a MOOC, or in a context sufficiently similar to be of interest to MOOC researchers; (c) which meet basic standards for quality research, including peer-reviewed work which contains sufficient description of their methods as to provide insight into the data and feature engineering, modeling, and experimental results. When a work was considered borderline on one or more of these criteria, we generally erred on the side of inclusion if it made a novel or relevant contribution to the literature. The literature surveyed was drawn from several top conferences and journals in the fields of learning analytics and educational data mining, computer science, web usage mining, and education, but was also collected from other sources (online searches, citations from other works surveyed).

We conducted a broad survey of existing research, hoping to unify work from a wide variety of disciplines which can be broadly considered predictive models. We evaluate work which studies environments that meet the definition of MOOCs described in Section \ref{sec:mooc-intro} above. Where studies are excluded, it is typically because they did not evaluate what we considered to be MOOCs, or did not meet other criteria discussed in Section \ref{sec:lit-review}. 

Several keywords were used to search prominent peer-reviewed conference, journal, and workshop proceedings in the fields of Learning Analytics and Educational Data Mining, including the Journal of Learning Analytics, the International Conference on Learning Analytics and Knowledge (LAK), Journal of Educational Data Mining , the International Conference on Educational Data Mining, the International Conference on Learning@Scale, the International Conference on Artificial Intelligence in Education, and the Journal of Artificial Intelligence in Education. Keywords used included ``MOOC'', ``predict'', ``model'', and ``dropout''. Additionally, we used the works cited in those works uncovered in our initial survey to ensure that we collected relevant work from the many other fields which have contributed research to predictive modeling in education, such as computer science, data mining, psychology, and educational theory. This review surveys work published in the year 2017 or earlier.

We note that select studies were still included despite not meeting individual components of this definition (for example, we do consider some work evaluating for-credit courses); in these cases we typically include such work either (a) for completeness due to the novelty or important contribution of the work, or (b) in order to err on the side of inclusion when the context of the course(s) under evaluation was not clear. We also note that the xMOOC phenomenon is not represented in our analysis, but this is in part because we were not able to identify any instances of xMOOCs being used with predictive student success models.

In particular, we also note that some work included in the review below might not have prediction as its stated aim. We believe, however, several such works are relevant to this review. ``Predictive'' modeling and modeling for data understanding are, as we discuss in Section \ref{sec:two-cultures}, two sides of the same coin -- both use statistical models of the data, which must capture relevant attributes and learn their relationship to an outcome of interest. While one work might construct a logistic regression model, for example, with the aim of understanding its parameters \citep[e.g.][]{Kizilcec2015-gu}, another might use the same modeling technique for a purely predictive goal \citep[e.g.][]{Whitehill2017-tt}. As such, techniques which are effective for one approach are often enlightening for the other as well. This is one of the core tensions of the ``two cultures'' -- while black-box models often fit the data better, they are more difficult to interpret; while data models are often highly interpretable, they are often so at the cost of the quality of fit. We thus found that many experiments which might not have the stated aim of prediction were still of great interest to readers of this review. We do, however, attempt to distinguish between works which are purely exploratory or descriptive (where the stated goal is not predictive) throughout the survey that follows.

\subsection{Activity-Based Models}

\textbf{Activity-based models} use behavioral data, evaluate behavioral outcomes, or are grounded in theories of learner behavior for predictive modeling. 

As shown in Figure \ref{fig:model_categories}, models utilizing activity-based features and outcomes are overwhelmingly the most common in the work surveyed. This is so for several reasons: first, as demonstrated in Figure \ref{fig:outcomes}, most of the works surveyed predict an engagement-based outcome related to dropout or course persistence. Activity features seem most appropriate for this type of prediction task (although more diverse feature sets may improve the quality and robustness of these models). Second, activity data are the most abundant and granular data available from MOOC platforms. Clickstream files (shown in Figure \ref{fig:clickstream-sample}) provide detailed interaction-level data about users' engagement with the platform, and such granular data is simply not available for any of the other model categories we survey. Collecting a similar level of granularity for these other feature types would require far more sophisticated data collection practices, such as affect detectors or other sensors, which are impractical at MOOC scale. Third, activity features appear to provide reasonable predictive performance even in non-activity-based prediction tasks, such as in grade prediction (e.g., \cite{Brinton2015-bx}). Indeed, it is reasonable to expect behavior to be associated with non-behavioral outcomes (i.e., learning). However, we note that state-of-the-art predictive models generally combine feature types to achieve a complete, multidimensional view of learners \citep[e.g.][]{Taylor2014-hu}.

The level of sophistication of the activity-based features in the works surveyed varies substantially, ranging from simple counting-based features \citep[e.g.][]{Kloft2014-kb, Xing2016-le} to more complex features, including temporal indicators of increase/decrease \citep{Veeramachaneni2014-ug, Chen2017-sf, Bote-Lorenzo2017-yh}, sequences \citep{Balakrishnan2013-ty, Fei2015-ea}, and latent variable models \citep{Sinha2014-lw, Ramesh2013-vu, Ramesh2014-cy, Qiu2016-ct}. Despite this variation, each of these typically uses the same underlying data source (clickstream, or a relational database consisting of extracted time-stamped clickstream events) and draws from a relatively small and consistent set of base features, including:
\begin{itemize}
    \item \textit{page viewing}, or visiting various course pages, such as video lecture viewing pages, assignment pages, or course progress pages; 
    \item \textit{video interactions},  such as play/pause/skip/change speed; 
    \item \textit{forum posting} or forum viewing (a more specific subset of page viewing which has received particular attention);
    \item \textit{content interactions}, which can take a variety of forms depending on the course and which may include assignment attempts, programming activity, peer assignment review, or exam activity.
\end{itemize} 

The relative consistency of the underlying activity-based feature sets and the few categories into which they can be distilled is largely a reflection of the consistency of the affordances available across the dominant MOOC platforms, particularly edX and Coursera: page viewing, video viewing, forum posting, and assignment submission were, until the introduction of relatively recent features such as interactive programming exercises, some of the only activities available to users of the platform, and the only activities recorded in Coursera clickstreams \citep{Coursera2013-jh}.

\subsubsection{Counting-Based Activity Features}

\cite{Kloft2014-kb} provided a foundational early predictive model, utilizing a Support Vector Machine (SVM) built on simple counting-based features extracted entirely from clickstream events. They find that high-level features related to activity (number of sessions, number of active days) were predictive of dropout during the first third of the course; measures of content interaction (wiki page views and homework submission page views) became more predictive in the middle third of the course; and navigation and general activity (number of requests, number of page views) were most predictive during the final third of the course. Kloft et al. use principal component analysis to demonstrate that successively generating a wider feature space by concatenating feature vectors for each subsequent week improves separability between dropout and non-dropout students by the final third of the course. This feature appending strategy has been widely adopted in predictive modeling in MOOCs \citep[e.g.][]{Xing2016-le}, likely due to the nearly universal structuring of MOOCs into weekly modules. \cite{Kloft2014-kb} only report the accuracy of their method, but based on the data they provide, the model's predictions offer less than a 5\% improvement on a majority-class prediction over the first 10 weeks of the course, when most dropouts occur (the challenge of evaluating this particular result using only accuracy highlights issues related to model evaluation and the lack of consistent metrics for reporting predictive results we discuss in Section \ref{sec:meth-res-gaps}). If predictive models are to be used to support \textit{early} interventions in MOOCs, more accurate predictions are required.

Kloft et al's results reinforce earlier findings from other digital education environments, such as \cite{Ramos2008-th}, who argued in 2008 in the title of their work that \textit{``Hits'' (not ``Discussion Posts'') predict student success in online courses}. This pre-MOOC study, conducted on an online university course, is notable for its finding that a simple count of page hits predicted student success (as measured by final course grade) better than either discussion posts or quiz scores, predicting between 7\% and 26\% of the variance in course grades. This finding has been reinforced in subsequent experiments evaluating behavioral features against other feature types in MOOCs \citep{Crossley2016-ij, Gardner2018-ga}.

\subsubsection{Models Utilizing Early Course Activity}

Several works attempt to address the need for \textit{early} predictions of student success in a MOOC. \cite{Jiang2014-vk} offers a simple logistic regression classifier based only on week 1 behavior which effectively predicts certification in a MOOC offered to university students. This model uses only four predictors representing different aspects of student engagement in week 1 of the course (average quiz score, number of peer assessments completed, social network degree, and an indicator for being an incoming university student at the institution offering the course), again suggesting that a limited, but diverse, feature space can effectively predict MOOC student success. However, the fact that this model was trained and tested on only a single MOOC -- one which may be particularly unique, because it was offered with an incentive (early enrollment in a university biology major) to current or prospective students at the hosting institution -- means that further replication is needed to determine the extent to which these results are generalizable. We highlight similar issues with comparing different experimental populations in MOOC research in Section \ref{sec:meth-res-gaps}.

Other work which attempts to perform more fine-grained dropout prediction with the intention of performing early intervention includes \cite{Xing2016-le}, which uses an ensemble of C4.5 tree and Bayesian Network models built on a set of counting-based engagement features and a Principle Component Analysis-based approach similar to \cite{Kloft2014-kb}. \cite{Baker2015-uz} find that early access to course resources in an e-learning history course, including a course textbook and its integrated formative assessments, provides accurate predictions of success or failure within the first two weeks of the course. In a pair of works which use more sophisticated temporal features to aid in early dropout prediction, \cite{Ye2014-st} and \cite{Ye2015-uf} find that fine-grained features related to either (a) temporal engagement with lecture quizzes or (b) the quantity of engagement with lecture quizzes improve models, but that once either (a) or (b) is included, adding the other provides no further performance gains. This result may suggest a plateau to the effectiveness of the features they evaluate, or it may highlight the need for more flexible modeling techniques to learn the complex patterns in rich, granular feature sets. 

\cite{Stein2014-kx} evaluate learner behavior in a microeconomics MOOC. Stein and Allione find that early engagement -- completing a quiz or a peer assessment exercise in the first week of a nine-week course -- is a significant predictor of persistence in the course, even when controlling for other behaviors. They conclude that ``the attrition pattern is not uniform among all enrollees, but rather there are distinct sub-groups of participants who reveal their type early on'' \citep[][p. 2]{Stein2014-kx}). This suggests that students' behavior early in the course might be particularly predictive of their final performance, which is a useful result for researchers or other stakeholders interested in obtaining accurate, early performance predictions.

A practical issue with ``early warning'' systems is that their predictions can change dramatically during the early stages of a course as the model predicts based on only small amounts of data. \cite{He2015-ab} address this challenge by using a smoothed logistic regression model trained from a previous offering of a MOOC to make calibrated predictions on a future offering which where fluctuation of predicted dropout probabilities over time is minimized. This smoothing provides stable predictions of at-risk students for early intervention, a useful property for real-world implementation which allows the students tagged as ``at-risk'' to remain relatively stable over time.

\subsubsection{Temporal and Sequential Activity Models}

An early approach to utilizing the \textit{temporal} nature of activity data (by using a model which captures transition probabilities over time from a weekly feature set) is \cite{Balakrishnan2013-ty}. This work uses a relatively small set of  features (cumulative percentage of available lecture videos watched,  number of threads viewed on the forum, number of posts made on the forum, number of times the course progress page was checked), compiled over each week of the course, to construct a Hidden Markov Model (HMM) to predict dropout. A particularly novel aspect of this work is the use of students' checking of their course progress page as an input feature. A challenge present in all predictive models of student success in MOOCs is accounting for students' diverse intentions (browsing, learning, completing, etc.). Balakrishnan and Koetzee introduce the course progress checking feature as an observable -- and effective -- proxy for an intention to complete: students who never check their course progress have a dropout rate of 20-40\% at each week of the course, while students who check their progress four or more times have a dropout rate of less than 5\% each week. This particular result suggests that finding observable proxies for student intentions is a tractable and useful problem for predictive models of student success in MOOCs.

In contrast to the simple feature appending approach used by e.g. \cite{Kloft2014-kb}, which shows variable (and only slight) improvement over weeks as data accumulates, more sophisticated temporal modeling approaches have demonstrated the ability to improve predictions more rapidly and consistently. \cite{Brooks2015-ej} examine how a higher-order time series method improves by exploring its incremental changes in performance with each additional day of MOOC data; they demonstrate rapid performance gains over the first three weeks of each MOOC evaluated. \cite{Fei2015-ea} explore sequential models, including a Long Short-Term Memory neural network (LSTM), which takes sequences of weekly activity feature vectors as its input. Fei and Yeung demonstrate consistent improvement in these models' performance as additional data is collected over course weeks, particularly over the initial weeks of a course. This model is directly compared to several others, outperforming (1) a Support Vector Machine (SVM; for reference to \cite{Kloft2014-kb} but with a different basis kernel); (2) two variants of Input-Output Hidden Markov Models (IOHMM; for reference to \cite{Balakrishnan2013-ty}, which uses a different HMM variant); and (3) logistic regression (compare to \cite{Jiang2014-vk, Veeramachaneni2014-ug, Liang2016-us}). 

The use of an LSTM by \cite{Fei2015-ea} is a promising approach, but further replication across a larger sample of courses is needed. This work also demonstrates how challenging it can be to compare results across machine learned models when exact replication of experimental populations and method is not possible (for example, Fei and Yeung cannot compare their model by using the data from \cite{Balakrishnan2013-ty}, nor can they perfectly reproduce the HMM model implementation from only the published description; we discuss this issue in Section \ref{sec:meth-res-gaps}), but their effort to provide these reference points is still useful. 

Additionally, \cite{Fei2015-ea} implement their model using three different definitions of dropout, which demonstrates the challenges of comparing predictive models of student success using published results (which often only vaguely describe outcome or feature definitions) and also suggests the robustness of their results. \cite{Wang2016-yc} evaluate a Nonlinear State Space Model (NSSM) in comparison to several other models, including an LSTM, and suggest that the NSSM achieves superior performance. We discuss the need for further comparative work in Section \ref{sec:future-research}.

Furthermore, we note that LSTMs and any deep neural network architectures require a large amount of data in order to accurately estimate the large number of model parameters involved. As a result, the use of these models is only available when large sets of training data (thousands or millions of instances) are available. This also points to the need for large, shared benchmarking datasets in the educational predictive modeling community, such as those provided by the MOOC Replication Framework (MORF) \citep{Gardner2018-hj} \footnote{\url{educational-technology-collective.github.io/morf/}} and DataStage\footnote{\url{https://datastage.stanford.edu/}}.

\cite{Sinha2014-sr} use sequential activity features in combination with higher-order graphical features (which represent the richness, repetition, and activity/passivity of students' interaction sequences) to predict dropout. They also conduct the useful comparison of whether using features from the current week only vs. a students' entire history improves performance, finding that the full history does not provide a significant improvement over current week only features. This result conflicts with \cite{Xing2016-le}, which finds that historical features improve the quality and stability of predictions in a single course offered on Canvas, but Sinha et al. use a larger and perhaps more representative sample of MOOCs.

\subsubsection{Latent Variable Modeling}

Latent variable modeling has been commonly applied to predictive models of student success, because of its ability to infer complex relationships between predictors in a data-driven way.

\cite{Ramesh2013-vu} apply Probabilistic Soft Logic (PSL) to a set of activity- and natural language-based features to model student performance. This work uses an expert-generated latent variable approach in which engagement is ``modeled as a complex interaction of behavioral, linguistic and social cues'' (p. 6). However, this particular method presents a potential barrier to practical implementation by utilizing only human-generated PSL rules. This is problematic for two reasons: (a) even experts may not be able to exhaustively identify the factors important to student success in MOOCs, particularly in a new course or a different domain (indeed, this is what motivates much of the work surveyed here), and (b) learning these features is itself the goal of data-driven predictive modeling. Manually defining latent engagement categories prevents truly data-driven discovery of latent user profiles or engagement types. Furthermore, by restricting the model to a small set of 5-7 features, this approach limits experimenters from learning about broader feature sets and their relationship to student success. \cite{Ramesh2014-cy} expands on their approach by using the latent variable assignments from this PSL method as predictors in a survival model.

In a pair of works utilizing the same underlying feature set, \cite{Halawa2014-pj} and  \cite{Kizilcec2015-gu} explore the use of learner activity features for predicting dropout in MOOCs. In the first of these works,  \cite{Halawa2014-pj} use a simple thresholding model to explore the use of counting-based learner activity features to predict dropout, theorizing that both observable learner activity and dropout are driven by latent, unobservable ``persistence factors'' which students possess to varying degrees. Halawa et al. show that this model is able to spot risk signals at least 2 weeks before dropout for over 60\% of the students in their experimental population (students who joined in the first 10 days of the course and have viewed at least one video), suggesting that early dropout prediction may be tractable for this group. \cite{Kizilcec2015-gu} applies this analysis to a sample of 20 MOOCs, utilizing the same feature set with a simple logistic regression model with similar findings.

\subsubsection{Course Metadata}

There has been a limited amount of prior work on studying aspects of courses themselves which may be relevant to student activity within the courses. In a work notable for its comprehensive sample of MOOCs, \cite{Evans2016-gj} examine a sample of 44 MOOCs and over 2 million learners, evaluating both student and course traits for association with engagement and persistence. Four findings are particularly relevant to student success prediction in MOOCs. First, early engagement (such as registering more than four weeks prior to course opening, or completing a pre-course survey) is the strongest predictor of completion. Second, the steep dropoff in engagement is ``very strong and nearly universal'' across the courses examined, which provides evidence supporting the implicit assumption of generalizability across courses in many other works. Third, the \textit{title} of individual lectures are associated with differing levels of engagement, with titles containing the words ``intro,'' ``overview,'' and ``welcome'' having significantly higher rates of watching. Fourth, the first offering of a course has significantly higher rates of completion than subsequent offerings -- an important finding with implications for the real-world deployment of models learned on data from previous courses.
 
 Additionally, \cite{Qiu2016-ct} evaluates the ways in which course subject interacts with learner demographics (i.e., gender) in predictive models. Qiu et al. find significant differences between the behavior of students in science MOOCs vs. non-science MOOCs. However, \cite{Whitehill2017-tt} find that models trained on data from many different domains are actually \textit{more} accurate than models trained on courses from only the same field as the target course, so perhaps these cross-disciplinary differences in student behavior can be addressed by using sufficiently diverse training sets to construct student models.

\subsubsection{Higher-Order Activity-Based Features}~\label{sec:higher-order}

Other work, utilizing more complex feature types, has also begun to emerge in MOOC research. This includes explorations of higher-order $n$-gram representations of learner activity data, which has demonstrated promising predictive performance \citep[e.g.][]{Brooks2015-fh, Brooks2015-ej, Li2017-yd}. In activity-based $n$-gram models, features are assembled using counts of unique sequences of events or behaviors; these features are then used to construct supervised learning models. This allows for the construction of large feature spaces which capture complex temporal patterns, and the frequency with which they occur. These works operate under the (often explicit) assumptions that \textit{sequences} of behavior, irrespective of the time gaps between them, contain richer information than individual events or counts of these events without considering the context of other neighboring events in time.

As discussed above, \cite{Sinha2014-sr} use $n$-gram features with a graphical model, and demonstrate that they can achieve reasonable predictive accuracy with only a single week of historical data.

We previously discussed \cite{Coleman2015-cq}, which applies topic modeling to sequences of learner data to learn ``profiles'' of MOOC learners based on their activity sequences (``shopping'', ``disengaging'', and ``completing'') . Each of these works and other sequence-based approaches discussed above (i.e., \cite{Balakrishnan2013-ty, Wang2016-yc}) can be thought of as capturing a temporal element of MOOC data. We argue in Section \ref{sec:temporal-modeling} below that further work in this vein is needed.

As we discuss below, feature engineering (not predictive modeling algorithms) is the primary driver of improvements in predictive modeling in MOOCs to date; future work should continue to pursue higher-order or other unique feature engineering approaches which capture information relevant to student success.

\subsubsection{Novel Feature Extraction and Prediction Architectures}

In a series of works, \cite{Veeramachaneni2014-ug}; \cite{Taylor2014-hu}; \cite{Taylor2014-pv}; and \cite{Boyer2015-lo} further demonstrate both the utility of effective feature engineering and how, when combined with effective statistical models, such methods yield performant student success predictors. These works use a combination of crowd-sourced feature extraction, automatic model tuning, and transfer learning to demonstrate several novel approaches to constructing activity-based models of student success in MOOCs.

\cite{Veeramachaneni2014-ug} use crowd-sourced feature extraction, leveraging members of a MOOC to apply their human expertise and domain knowledge to define behavioral features for stopout prediction. The authors find that these crowd-proposed features are more complex and have better predictive performance than simpler author-proposed features for all four cohorts evaluated (passive collaborator, wiki contributor, forum contributor, and fully collaborative). This work utilizes a simple regularized logistic regression for the predictive model, again demonstrating that many effective predictive models of student success in MOOCs have relied on clever feature engineering, not sophisticated algorithms. The use of regularization common in MOOC research (see \ref{sec:synthesis} for details) due to the large number of correlated predictors often present in student models.

\cite{Taylor2014-hu} applies the feature set from \cite{Veeramachaneni2014-ug} to explore over 70,000 models using a self-optimizing machine learning system. However, the consideration of such a massive model space on only a single cohort of students virtually guarantees at least \textit{some} success in prediction due to chance alone. Further validation and testing of the ``best'' models identified in this work are needed. In many ways, this work is an extreme example of a common approach where large model spaces are explored without utilizing effective statistical evaluation methods, resulting in performance data whose significance and generalizability is difficult to interpret.\footnote{We discuss concerns related to large numbers of comparisons, including with self-optimizing or auto-tuning machine learning toolkits, in a forthcoming work.}

\cite{Boyer2015-lo} explore transfer learning using a subset of the feature set from these prior works. \cite{Boyer2015-lo} is notable for its experimental treatment of how previous iterations of a MOOC can be used to predict on future iterations, which is how such models are used in practice. This setup addresses one challenge of model deployment in ``live'' courses, and provides initial data on effective transfer architectures for doing so. While many of the experimental results are inconclusive, Boyer and Veeramachaneni demonstrate two particularly important findings.

First, Boyer and Veeramachaneni find that \textit{a posteriori} models -- built retrospectively using the labeled data from the target course itself, which is the dominant experimental architecture used across our survey -- presents ``an optimistic estimate,'' and that such models ``struggle to achieve the same performance when transferred'' (we discuss potential issues with \textit{a posteriori} models, and their prevalence across the work reviewed, in Section \ref{sec:realistic-experiments}). They conclude: ``when developing \textit{stopout} models for MOOCs for real time use, one \textit{must} evaluate the performance of the model on successive offerings and report its performance'' (emphasis from original) \citep[][p. 8]{Boyer2015-lo}. This and other work \citep[e.g.][]{Brooks2015-ej, Evans2016-gj, Whitehill2017-tt} suggests that there is a great deal of work to do in replicating, re-evaluating, and exploring the generalizability of previous stopout prediction work performed using an \textit{a posteriori} architecture. For one example of work which compares models evaluated both within and across courses, see \cite{Wang2016-yc}, which presents evidence that the ``penalty'' for model transfer across courses might be minimal.

Second, Boyer and Veeramachaneni find that an \textit{in situ} prediction architecture transfers well, achieving performance comparable to a model which considers a users' entire history (which is not actually possible to obtain during an in-progress course). \textit{In situ} architectures consider data and proxy labels from the same course to train a model (rather than true labels of future stopout, which are not known at the time of training/prediction in this realistic formulation of the task). This finding presents a possible approach to resolve the problems with using \textit{a posteriori} modeling in practice, and is supported by other work \citep[e.g.][]{Whitehill2017-tt}.

In a different examination of model transfer, we surveyed two works \citep{Vitiello2017-au, Cocea2007-eu} which examine how models trained on one \textit{platform} transfer to another (the former studies a MOOC environment; the latter a web-based e-Learning system). Both demonstrate that high-performing features are stable even for models trained across different platforms. This suggests that effective activity-based feature sets may transfer well across MOOC platforms (when the data they require is available from these platforms), but further research is required to verify this result.

Another innovative approach to representing and modeling activity sequences is presented in \cite{Zafra2012-qa}, where a multi-instance genetic algorithm is used to model ``bags'' of instances representing information about each students' activity across various behavior and resource types. This algorithm is particularly unique in its ability to resolve missing-data issues with sparse features (such as forum posts) available only for a small subset of learners \citep{Gardner2018-ga}: the multi-instance algorithm accepts bags of varying sizes to accommodate the unique subsets of activities displayed by each student. Zafra and Ventura's experiment is conducted in the context of a set of e-learning courses offered via Moodle, but the authors argue that this approach is scalable and that it would be particularly useful for large online courses due to the heterogeneous student behavior patterns in these courses.

\subsection{Discussion Forum and Text-Based Models}~\label{sec:text-lang-models}

\textbf{Discussion Forum and Text-Based Models} use natural language data generated by learners and/or use linguistic theory as the basis of student models. 

Threaded discussion fora are a prominent feature of every major MOOC platform and are widely used in most courses. Detailed analysis of the data from discussion fora provides the opportunity to study several dimensions of learner experience and engagement which are not detectable elsewhere. This includes a rich set of linguistic (measured by analysis of the textual content of forum posts), social (measured by the networks of posts and responses, or actions such as up/downvotes), and behavioral features not available purely from the evaluation of clickstream data. \cite{Gardner2018-ga} argues that understanding the individual contributions that separate data sources make to predictive models is useful in determining whether scarce developer time ought to be dedicated to feature engineering, extraction, and modeling from those sources. This is particularly relevant to the complex data in discussion fora: extracting the features required to construct many of the models surveyed below can be time- and developer-intensive; it should only be done if the benefits (in terms of improved prediction or insight) justify these costs.

A foundational series of forum-based predictive work is that of Ros\'e, Wen, Yang, and collaborators \citep{Rose2014-mk, Wen2014-xr,Yang2015-gy}, and particularly \cite{Yang2013-zq}. This series of work uses discussion forum data to identify the social environmental characteristics that are most conducive to persistence or sustained engagement in a MOOC. \cite{Yang2013-zq} uses forum post data to explore the predictiveness of three types of features for forum posters in a single MOOC: cohort (the week in which a user joined the course), forum post (threads started, post length, content length), and social network (several metrics, including centrality, degree, authority, etc.). Of 16 variables considered in a variety of model specifications, Yang et al. find only three that are significant predictors for these students: being a member of cohort 1 (joining in the first week of the course), writing forum posts that are longer than average, and having a higher than average authority score are all associated with a lower probability of dropout. \cite{Rose2014-mk} adds subcommunity membership to this feature set; in this case, cohort 1 membership is still significant, but their finding on authority is \textit{reversed} -- with a ``nearly 100\% likelihood of dropout on the next time point for students who have an authority score on a week that is a standard deviation larger than average in comparison with students who have an average authority score'' (p. 198). A Mixed Membership Stochastic Blockmodel (MMSB) is used to identify the subcommunities utilized as predictors. These results suggest that the social factors, and not the language, of discussion fora may be more effective predictors of dropout for students who post in the fora than the text of the post itself. Work by \cite{Wen2014-xr} and \cite{Yang2015-gy} are discussed in Section \ref{sec:cog-models} below.

\cite{Robinson2016-yr} apply natural language processing to pre-course open-response questions on learners' anticipated utility of course material. Using unigram features improves dropout prediction over a demographics-only model for students intending to complete the course. A series of richer features from the Linguistic Inquiry and Word Count (LIWC) framework \citep{Pennebaker2015-nr} are not found to be significant predictors of dropout. However, the final model in this work achieves a relatively low AUC (59.8) despite being evaluated using a \textit{post hoc} architecture (cross-validated testing using the same course on which the model was trained) and analyzing a subpopulation which is less than 5\% of the students who registered for the course, and less than 7\% of the students who engaged with the course during the first two weeks. This suggests that the model is not particularly well fit to the data, even given the small subsample of the course population used; further research would help identify the extent to which these results generalize to other populations. We note in Section \ref{sec:two-cultures} that the lack of fit from using simple, but interpretable, data models is an argument in favor of more complex (but less interpretable) models; we expect future work to continue the trend of bridging this fidelity-interpretability gap.

\cite{Dowell2015-ip} explore discourse features generated from forum posts, which are able to account for 5\% of the variance in learner final grades (in contrast, a model with discourse features and participant features explains 93\% of this variance). For the most active students (the top quartile, based on count of posts), discourse features explain 23\% of the variance in performance. The authors conclude that discourse features are most effective at predicting performance for the most active students. Considering that forum posters might already be considered the most active and engaged students in a MOOC, these results suggest that the predictive usefulness of discourse analysis might be limited to a small subpopulation of learners in many MOOCs.

\cite{Crossley2016-ij} compares the predictiveness of clickstream-based activity features and natural language processing features. They find that clickstream-based activity features are the strongest predictors of completion, but that NLP features were also predictive; the addition of clickstream-based activity features improves the performance over a linguistic-only model by about 10\% \citep{Crossley2016-ij}. While the sample size in this experiment is only the small subset of students who both posted in a forum and completed an assignment, it makes a useful and important contribution to the literature by systematically comparing two of the dominant feature sets (activity and forum features).  Further exploration, including systematic, statistical evaluation of the predictive efficacy of each feature set across larger course populations, is needed to validate these results and explore the degree to which they generalize.

\cite{Tucker2014-ue} investigate the correlation between students' sentiment in posts about specific assignments and their performance on those assignments in an art MOOC, finding a modest negative correlation. They also find a modest positive trend in forum post sentiment over the duration of the course. Other predictive work related to sentiment analysis includes \citep[e.g.][]{Wen2014-pv}, which demonstrates an association between sentiment and attrition which appears to differ by course topic, and \cite{Chaplot2015-dk}.

\cite{Adamopoulos2013-pv} presents an alternative approach to using text analysis to understand student success in MOOCs by analyzing public student reviews of MOOCs. Adamopoulos suggests that student course completion is influenced by perceived course quality, course characteristics (topic, perceived difficulty), characteristics of the offering institution (e.g. university ranking or prestige), platform characteristics (i.e. usability), and student characteristics (i.e. gender). This work matches other examinations of factors affecting student dropout in e-learning and distance learning courses  \citep[i.e.][]{Levy2007-ai, Ji-Hye_Park2009-sq}.

We conclude this section with a brief note. One of the particular challenges of working with text and forum data in MOOCs is the relative sparsity of this data: as optional activities, forum posts (as well as up- and down-voting, pre- and post-course surveys, and other questionnaires) are only available for the subpopulations which elect to participate in them. In most cases, this is a fraction of the population; sometimes as little as 5\% (see Table \ref{tab:subpopulations}). Therefore, work which utilizes these data sources typically restricts its experimental population only to the small subpopulation of students for whom this data is available. While this can still lead to interesting and informative insights about this subgroup, we believe that work which excludes 95\% of the participants in a course ought to be considered either exploratory or very limited in its scope. This observation applies to a great deal of MOOC research, as we discuss in Section \ref{sec:meth-res-gaps} below, but it is particularly problematic (and is least often acknowledged) in language-based experiments.

\subsection{Social Models}\label{sec:social-models}

\textbf{Social models} use observed or inferred social relationships, or theories of social interaction, as the foundation for student models. 

Many works surveyed use discussion fora to construct social networks where students are nodes and various reply relationships constitute edges. For example, \cite{Joksimovic2016-mp} uses two sessions of a programming MOOC, offered in English and Spanish, respectively, to evaluate the relationship between social network ties and performance (specifically, non-completion vs. completion or completion with distinction). Students who achieved a certificate or distinction were more likely to interact with each other than with non-completers (in contrast, \cite{Jiang2014-iv} find in a different set of MOOCs that learners tend to communicate with others in \textit{different} performance group). Furthermore, \cite{Joksimovic2016-mp} find that weighted degree centrality was a statistically significant predictor of completion with distinction in both courses, and a significant predictor of basic completion in the Spanish-language course, while closeness and betweenness centrality showed more variable and inconsistent effects across courses. They conclude that structural centrality in the network appears to be positively associated with course completion \citep{Joksimovic2016-mp}. The finding matches that of \cite{Russo2005-nw}, who also identified centrality as a statistically significant predictor of student performance in a small e-learning course. In a related work, \cite{Dowell2015-ip} evaluate how social centrality itself can be predicted by text discourse features\footnote{While the current survey is not specifically interested in the prediction of these outcomes, we include these works on the basis that they contain other, more direct predictions of student success in MOOCs or generate insights relevant to such predictions.}, finding that discourse features explain about 10\% of the variance in performance (compared to 92\% explained with a model using discourse + participant features); this increased to 23\% explained for the most active participants in the fora.

\cite{Yang2014-ka}, also discussed above, use a graph clustering method to construct probabilistic models of students' social network membership over the subcommunities in a course. Membership in some subcommunities defined by the MMSB are significantly predictive of dropout, while others are not; the number of subcommunities that are significant predictors varies between two and four across their three-MOOC sample (the authors consider up to 20 subcommunities per course). Other work has identified social networks as effective predictors of student performance in traditional academic courses \citep{Fire2012-ye, Gasevic2013-rq}.

\cite{Agudo-Peregrina2014-zg} examines social \textit{interactions} in online courses, finding that student-student, student-teacher, and student-resource interactions are all significantly related to learner performance, while the same interactions are not significant predictors in courses with an in-person component. Whie this work is not conducted in MOOCs, it demonstrates how broader elements of student engagement with other students and teachers might take on special importance in digital learning environments. 

More research on the impact of social networks in MOOCs, and further exploration of external social network data, is necessary. Social networks appear to be an important factor in students' learning, but are challenging to measure with existing MOOC data and even harder in relatively small, single-course samples. The use of external digital social networks (such as data from Facebook or LinkedIn) is rare in MOOCs, despite the richness of these data sources. Instead, existing research appears to be overly reliant on discussion fora as sources of social network data. The examination of novel data sources on social factors stands to substantially influence the research consensus in this area and would likely lead to novel and useful findings about the relationships between social connectedness and student success in MOOCs.

\subsection{Cognitive Models}\label{sec:cog-models}

\textbf{Cognitive models} use observed or inferred cognitive states, or rely on theories of cognition, as the basis for student models. 

While MOOCs are ultimately concerned with impacting learners' cognitive states (because learning is a cognitive process), surprisingly little research has attempted to explore the use of cognitive data in MOOCs. This may be, in part, because of the unique challenges of collecting this data, especially relative to the ease with which other rich data sources (activity, forum posts, etc.) can be collected from MOOC participants. A substantial portion of the work on cognitive states in MOOCs involves novel data collection methods, from biometric tracking \citep[e.g.][]{Xiao2015-me} to contemporaneous questionnaires \citep{Dillon2016-fa}.

\cite{Wang2015-oy} use discussion forum data to investigate ``the higher-order thinking behaviors demonstrated in student discourse and their connection with learning'' (p. 226). Hand-coded data, using a learning activity classification scheme from cognitive science research, is used to evaluate several learning outcomes. Of particular interest is the authors' finding that students who have demonstrated ``active'' and ``constructive'' behaviors in the discussion forum -- which demonstrate higher-level cognitive tasks such as synthesis, as opposed to merely paraphrasing or defining -- had significantly more learning gains than students who did not use these behaviors. This work demonstrates that useful cognitive data that is relevant to student performance can be extracted from discussion forum posts, even using relatively simple models (a bag-of-words and linear regression). Furthermore, it suggests that cognitive strategies -- if they can be effectively identified -- appear linked to student performance in MOOCs, and that cognitive theory can inform predictive models in MOOCs.

\cite{Wen2014-xr} and \cite{Yang2015-gy} extend their work discussed in Section \ref{sec:text-lang-models} work to use linguistic features of forum posts to identify the cognitive states they express; in particular, they seek to identify learner \textit{motivation} and the degree of \textit{confusion}. \cite{Wen2014-xr} use forum posts to derive (a) cognitive engagement features from the presence of unigrams in post text, and (b) human-coded learner motivation features. They find that these are significant predictors of dropout, using the survival modeling approach implemented in their previous work. \cite{Yang2015-gy} examines confusion in the text of forum posts, finding that the influence of confusion varies across courses, and that different types of confusion are significant predictors of dropout in each of the two courses evaluated. Yang et al. attribute this to differences in the domain of these two courses. 

\cite{Sinha2014-lw} uses activity data to infer cognitive states by generating an ``information processing index'' for each student based on an expert-generated taxonomy of user interaction sequences defining various behavioral actions (e.g. ``clear concept,'' ``slow watching,'' or ``checkback reference'') and weights which the authors manually assign to each action group. Again, however, using manually-defined features risks injecting experimenter bias into the model instead of generating truly data-driven features in this model. \cite{Sinha2014-sr} also uses interaction sequences to infer the presence of cognitive states; as mentioned above, this work attempts to discern, for example, the activity/passivity of a user based on the observed sequences of behaviors.

\textit{Emotions} are cognitive states which have received particular attention in MOOC modeling research. \cite{Dillon2016-fa} use self-reported emotional states to examine the relationship between emotions and activity type; co-occurring emotional states; and the relationship between emotions and dropout. Anxiety, confusion, frustration, and hope are each significantly correlated with dropout. Initial work by \citep[e.g.][discussed above]{Wen2014-pv, Chaplot2015-dk, Tucker2014-ue} utilizing sentiment analysis also suggest that information related to emotional states captured from learner-generated text can be useful in dropout prediction. \cite{Gutl2014-vx} evaluate learner emotions by administering questionnaires during learning activities, finding no significant difference between the relative proportion of happiness vs. sadness, anxiety, and anger between completers and non-completers. \cite{Russo2005-nw} explore whether network centrality and prestige can predict ``affective learning'' -- how students feel about a course -- in an e-learning course, but find that neither is a significant predictor. While affect has been studied in K-12 education and in digital cognitive tutoring environments \citep[e.g.][]{Pardos2013-hc}, there is comparatively less research on emotions in MOOCs. The use of information corresponding to emotional states represents a useful line of inquiry for future work.

\cite{Xiao2015-me} and \cite{Pham2015-cv} use heart rate tracking on mobile phones to conduct ``Implicit Cognitive States Inference,'' whereby MOOC learners' cognitive states (mind wandering and interest/confusion) are predicted from mobile phone measurements. This work is a proof-of-concept, but given the growth of both mobile devices for learning and the expansion of sensors and multimodal learning analytics, it points to potential future directions for student models that measure learners directly (not simply their navigation or submission behavior) and respond to real-time physiological, emotional, or cognitive feedback.

 \cite{Street2010-rz} reviews eight different studies of factors for student dropout of distance learning courses, with a focus on self-reported mindsets and attitudes which contribute to student success. Street finds that several internal factors (self-efficacy, self-determination, autonomy, and time management), external factors (family, organizational, and technical support), and course factors (relevance, design) all significantly impact learners’ decisions to persist or drop such courses. Other work surveying participants in e-learning courses finds similar influence of family support, organizational support, relevance, and other individual characteristics on individuals' decisions to drop out in this context \citep{Ji-Hye_Park2009-sq}. Although neither of these works can be considered predictive, they provide insight into cognitive factors which may be contributing to student outcomes in MOOCs.

\cite{Greene2015-no} explores students' perceived relevance of course material, commitment, and students' implicit theories of intelligence, as well as demographic indicators and information about students' prior experience with MOOCs. Self-reported commitment is reported as one of the strongest predictors of dropout, but students' implicit theories of intelligence are not strongly associated with dropout. They also find that intended hours spent on the MOOC is a significant predictor of exam scores, but that implicit theory of intelligence was not. We note that the relationship between intention and student success is reinforced in other work by \cite{Balakrishnan2013-ty}, which measured intention by students' views of course progress pages; \cite{Gutl2014-vx} finds a high level of self-reported motivation for both dropout and non-dropout students.

Much of the work in this section involves novel data collection methods. Similar to our findings on social factors above, there is a need for future research to move beyond questionnaires and self-reports as the sole source of cognitive data from learners. As sensing technology becomes increasingly affordable, and as users are increasingly already equipped with sensors inside their own devices (such as smartphones and tablets), the type of data required for this type of research should become increasingly accessible for researchers. There are many canonical cognitive findings in educational research which have yet to be explored or replicated in a MOOC context, and future work is needed to determine the limitations of these findings from traditional brick-and-mortar classrooms when applied to MOOCs.

\subsection{Learning-Based Models}

\textbf{Learning-based models} use observed student learning or performance on course assignments or theories of student learning as the basis for predictive modeling. 

While the formal purpose of a MOOC is, broadly construed, for participants to learn, the use of learning-based features and outcomes in predictive MOOC models has been surprisingly limited, as shown in Table \ref{tab:features-outcomes}. Much of the work in this section draws upon methods derived from the broader psychometrics, learning analytics, and educational data mining communities, applying well-known methods (e.g. Item Response Theory, Bayesian Knowledge Tracing) to MOOC data.

Several predictive studies in MOOCs discussed above attempted to predict learning-based \textit{outcomes}, despite being otherwise focused on different theoretical or modeling approaches. Some previously-discussed work in this category includes \citep{Brooks2015-ej, Greene2015-no, Kennedy2015-jw, Li2017-yd, Ye2014-st}. Such work predicts outcomes such as pass/fail, final grade, assignment, or exam prediction (further data on the use of learning-based prediction outcomes is provided in Figure \ref{fig:outcomes}).

\cite{Ren2016-tr} explore the use of ``personalized linear regression'' for predicting student quiz and homework grades, finding that this approach outperforms KT-IDEM, an item-level variant of Bayesian Knowledge Tracing widely researched in intelligent tutoring systems, in predicting homework scores across two MOOCs. 

\cite{Garman2010-td} applies pre-existing learning assessment to online courses by administering a commonly-used reading comprehension test (the Cloze Test) to students in an e-learning course. Garman finds that reading comprehension is positively associated with exam performance and overall course grade, but finds no association between reading comprehension and online open-book quizzes or projects. Garman argues that this is because online tasks are more under control of the student (taken independently, with fewer or no time constraints), while exams and course assignments occurred in an in-person environment with time constraints. While this study is administered in an e-learning course, these findings are relevant to MOOCs given the degree to which MOOC participants are expected to read and comprehend substantial amounts of text independently. \cite{Wojciechowski2005-du} also find significant relationships between student reading comprehension (as measured by ASSET scores and ACT English scores) in university e-learning courses.

\cite{Kennedy2015-jw} evaluate how prior knowledge and prior problem-solving abilities predict student performance in a discrete optimization MOOC with relatively high prior knowledge requirements, drawing on robust learning theory results from in-person courses. Prior content knowledge and problem solving abilities are measured using two performance tasks. The prior knowledge variables alone account for 83\% of the variance in students' performance in this MOOC. The relationship between prior knowledge and student performance is well-documented in traditional education research, but is largely unexplored in MOOCs, despite the potential presence of many more students who lack prerequisite prior knowledge in MOOCs relative to traditional higher education courses. Further research on both data collection (i.e., methods for efficiently measuring learners' prior knowledge at scale) and on the impact of prior knowledge on learner outcomes is a useful avenue for future research.

Time-on-task and task engagement are also student performance concepts which have been applied extensively to educational contexts outside of MOOCs. \cite{Champaign2014-zz} evaluate how learner time dedicated to various tasks within the MOOC platform (assignment problems, assessments, e-text, checkpoint questions) correlates with their learning gain and skill improvement in two engineering MOOCs. They find \textit{negative} correlations between time spent on a variety of instructional resources and both skill level and skill increase (i.e., improvement in students' individual rate of learning), using assessments calibrated according to Item Response Theory. Champaign et al. find these results ``obviously discouraging'' (p. 18), but their evaluation is purely correlational. They note that the observed association is likely due to struggling students spending more time working with learning activities. A more fine-grained analysis is needed to determine whether the results are truly causal, or perhaps instead indicative of other behavior, such as productive struggle. This work certainly suggests that further evaluation is necessary to measure whether students are truly learning in MOOCs (as opposed to high-skill students succeeding, while low-skill students drop out) and what types of resources and affordances best support learning. \cite{Cocea2007-eu} also address the task of evaluating students' engagement with content and explore the task of student engagement prediction in a web-based e-Learning system; this work demonstrates that accurate engagement predictions (based on expert-rated engagement) can be made using relatively simple activity features extracted from log files.

\cite{Koedinger2015-ef} examine the impact of using interactive educational resources in MOOCs versus using passive informational resources (videos, text) available in many MOOCs. Specifically, this work examines the use of interactive tools from the Open Learning Initiative, which were embedded into the Coursera platform. They find that learners using more interactive resources learn significantly more than those who read more text or watch more videos, estimating the impact of a 1-standard deviation increase in interactive resource use to be more than six times that of a 1-standard deviation increase in watching or reading. However, they find that the use of interactive resources was not a significant predictor of dropout, with quiz scores and quiz participation instead being significant predictors. This suggests that while these resources may indeed assist students in \textit{learning} more, this may not translate directly into course completion. This work highlights the importance of evaluating results along multiple outcome dimensions in MOOCs. 

\cite{DeBoer2014-mq} find that time spent on homework and labs in a Circuits and Electronics MOOC on edX predict higher achievement on assignments, while time spent on the discussion board or book is less predictive or not statistically significant. Additionally, time on the ungraded in-video quiz problems between lecture videos is found to be more predictive of achievement than time on lecture videos themselves.

Peer learning and peer assessment are also important theoretical concepts in education, but have seen only limited applications in MOOCs to date. \cite{Ashenafi2015-fo} and \cite{Ashenafi2016-df} examine models for student grade prediction which only use peer evaluation; these models are applied in traditional courses with web-based components but the authors argue that their findings are also applicable to MOOC contexts. Peer assessment is used extensively in MOOCs \citep{Jordan2015-op}  and its predictive capacity is largely unexplored. 

\cite{Brinton2015-ya} explore using platform clickstream data to build models of whether learners are Correct on First Attempt (CFA) in answering questions in a MOOC. After building models to predict CFA, these predictions are used as features in a model to predict students' future quiz performance. Brinton and Chiang demonstrate potential performance gains from this approach, suggesting that not only effective feature engineering, but also the predictions of intermediate models, can improve predictions of student success in MOOCs. \cite{Brinton2015-bx} extends this work with a sequence-based input approach. \cite{Sinha2015-yj} use a sequence-based approach to student learning, modeling the \textit{outcome} as a sequence and predicting sequences of student grades using Conditional Random Fields.

\cite{Kotsiantis2010-hs} apply various incremental algorithms to student performance prediction using a dataset of student grades in a distance education course. They find that an ensemble of incrementally-trained predictive models can achieve improved final exam pass/fail predictions over the base learners. While this model is applied to a single higher education distance learning course, it demonstrates a successful application of a technique -- incremental model training, requiring only a single pass through large datasets -- which may be particularly useful with the massive datasets in MOOCs. Further exploration of these techniques stands to make real-time training and prediction more tractable. \cite{Sanchez-Santillan2016-ml} also explores the use of incremental interaction classifiers using Moodle course data.

As the functionality of MOOC platforms and the associated tools used within those courses -- Integrated Development Environments (IDEs), notebook environments such as iPython and Jupyter, etc. -- have expanded, so too has the student performance and activity data available to instructors. Recent work has begun to evaluate this data. \cite{Hosseini2017-kp} uses a plugin in the NetBeans IDE to collect detailed data on student problem-solving in Java programming assignments to predict student problem-solving and learning in two programming courses and two MOOCs. The work evaluates both stereotype-based and fully data-driven models constructed using a ``genome'' representing student problem-solving behavior extracted from students' program submissions. Performance Factors Analysis is used to compare several models, and the authors identify clusters of students based on their problem-solving activity (``tinkerers'', ``movers''). While the authors uncover some apparent relationships between problem-solving behavior and learning, they conclude that there are both strong and weak students within each group -- these behavioral profiles are not, as constructed, predictive of learning. Hosseini et al. conclude that ``finding a useful learning-focused stereotype, like good students or slow students, is not trivial.  There might be students who approach learning differently, but the distinction between these approaches are orthogonal to the conventional dimensions that we apply to quantify learning'' (p. 83). This suggests that further evaluation of data-driven profiles of learning behavior are required in order to construct accurate models of how this behavior predicts student learning.

\subsection{Demographics-Based Models}~\label{sec:demographic-models}

\textbf{Demographics-Based Models} utilize learner attributes which remain static over the interval of a course to predict student success.

In this section, we explore work which student demographics to understand and predict student success in MOOCs. This work often utilizes optional surveys about learner demographics.

Several works have investigated the relationship between learner demographics and their success in MOOCs. Similar to research in more traditional educational contexts, the primary focus of this research is in understanding for which groups of students MOOCs may be more or less effective. In general, this work therefore tends toward explanatory or data modeling.

In an analysis of edX's first course, \cite{DeBoer2013-ki} find that having taken differential equations (a recommended prerequisite for the course), having a parent who is an engineer, and working with the teacher offline are significant predictors when controlling for other behavioral and academic factors; they do not find a relationship between gender and achievement for the survey completers examined. This finding regarding prior knowledge reinforces \cite{Kennedy2015-jw}, discussed previously. Similarly, \cite{Dupin-Bryant2004-bd} show that prior computer experience is also a predictor of retention in online distance education courses, likely because such students are better prepared to learn and engage with course content by computer.

\cite{Stein2014-kx}, discussed previously, evaluates a range of demographic factors for students who completed a pre-course survey, finding that self-reported motivation for taking the course is \textit{not} a significant predictor of completion, but that age is (with both young and very old students more likely to disengage).

\cite{Greene2015-no} also explores a combination of motivational factors (discussed in Section \ref{sec:cog-models} and demographic factors, finding that demographic variables including age, prior education, and prior experience with MOOCs are significant predictors of both dropout and achievement.

\cite{Qiu2016-ct} examines the impact of both gender and level of education on forum posting, total active time, and certification rate for a sample of XuetangX MOOCs. They find that being female is associated with higher rates of forum posting and replying, more time spent on video and assignment activity, and higher certification rate in non-science courses, while female is associated with each of these outcomes being \textit{lower} in science courses (only the association between female and forum replies and certification rates in science courses were not statistically significant at $\alpha = 0.1$). With respect to level of prior education, \cite{Qiu2016-ct} find that students with a bachelors degree ask more questions, particularly in non-science courses. They also report that students with a graduate degree are not as active as those with bachelors in terms of asking questions, but are instead more active in answering questions, particularly in science courses.

Specific findings related to various demographic features are multifarious, but comparing the magnitude of findings across different studies can be challenging when different controls are included in various models. \cite{Stein2014-kx} find that age is a significant predictor of MOOC completion. \cite{Greene2015-no} also find age, prior education, and prior experience with MOOCs to be significant predictors of both dropout and achievement. \cite{Reich2014-ob} finds that intention to complete is a stronger predictor than any of several demographic traits measured across a sample of 9 MOOCs. In a review of works surveying potential causes of MOOC dropout rates, \cite{Khalil2014-cw} also find that lack of time, lack of motivation, lack of interaction, and ``hidden costs'' (such as paid textbooks needed for reference, or paid certificates of which learners were unaware) contribute to MOOC dropout.

Several works have evaluated the predictiveness of demographics in e-learning or distance learning courses. While these are not directly analogous to MOOCs, the conclusions of such research can suggest useful starting points for further research into demographic and other factors which may contribute to MOOC dropout. For example, \cite{Willging2009-eu} use a post-course survey to understand explanatory factors underlying student dropout in an online human resources masters program. The authors find that demographics are not associated with dropout in the courses evaluated, and that reasons for dropout vary considerably by individual including personal reasons, job-related reasons, program-related reasons, and technology-related reasons.

\cite{Brooks2015-fh} examines whether demographics can improve the predictive performance of activity-based predictive models, showing that demographics ``have minimal predictive power when determining the academic achievement of learners enrolled in MOOCs.'' In particular, \cite{Brooks2015-fh} demonstrates that demographics-based models underperform activity-based models in MOOCs even early in the course when activity data is minimal, and that demographic features provide no discernable improvement over activity-only models (and actually degrade their performance in the second half of the course, as activity data accumulates). This stands in contrast to prior machine learning research in other educational domains, which suggests that demographics may be strong predictors of online course performance in traditional distance learning contexts \citep[e.g.][]{Kotsiantis2003-km}). The work of Brooks et al. highlight how the complexity and heterogeneity of MOOC learners require new and potentially more sophisticated student models, and how demographic findings may be less powerful than other unique, rich sources of data available in the contexts of MOOCs.



\section{Synthesis: Trends in Predictive Models of Student Success in MOOCs}\label{sec:synthesis}

In this section, we present high-level synthesis and conclusions from this survey of predictive modeling work in MOOCs to date, including data on the methods and findings of this work. We profile the data sources, methodologies, and experimental populations evaluated in these works. We find evidence that (a) a small number of MOOC platforms and raw data sources are used as the basis for the majority of MOOC research to date, and (b) a similarly dominant group of methodologies (activity-based features, tree-based and generalized linear modeling algorithms) that are used for these experiments. Together, these trends suggest a need for future research comparing these methods (particularly when considered in light of the many different success metrics used to evaluate these models across works surveyed), and exploring the use of other techniques and methods described in Section \ref{sec:future-research}. 

\subsection{Data Sources: Platforms and Raw Data Sources}

Little attention has been paid to the data sources used in predictive modeling research in MOOCs. Understanding which data sources are effective for prediction, and which are unexplored, provides a useful foundation for future work. Also, because feature extraction requires significant expense, both in terms of development time and computation time, recognizing which data sources are most useful can improve the efficiency of predictive modeling work in practice.

\begin{figure}
    \centering
    \includegraphics[width=\textwidth]{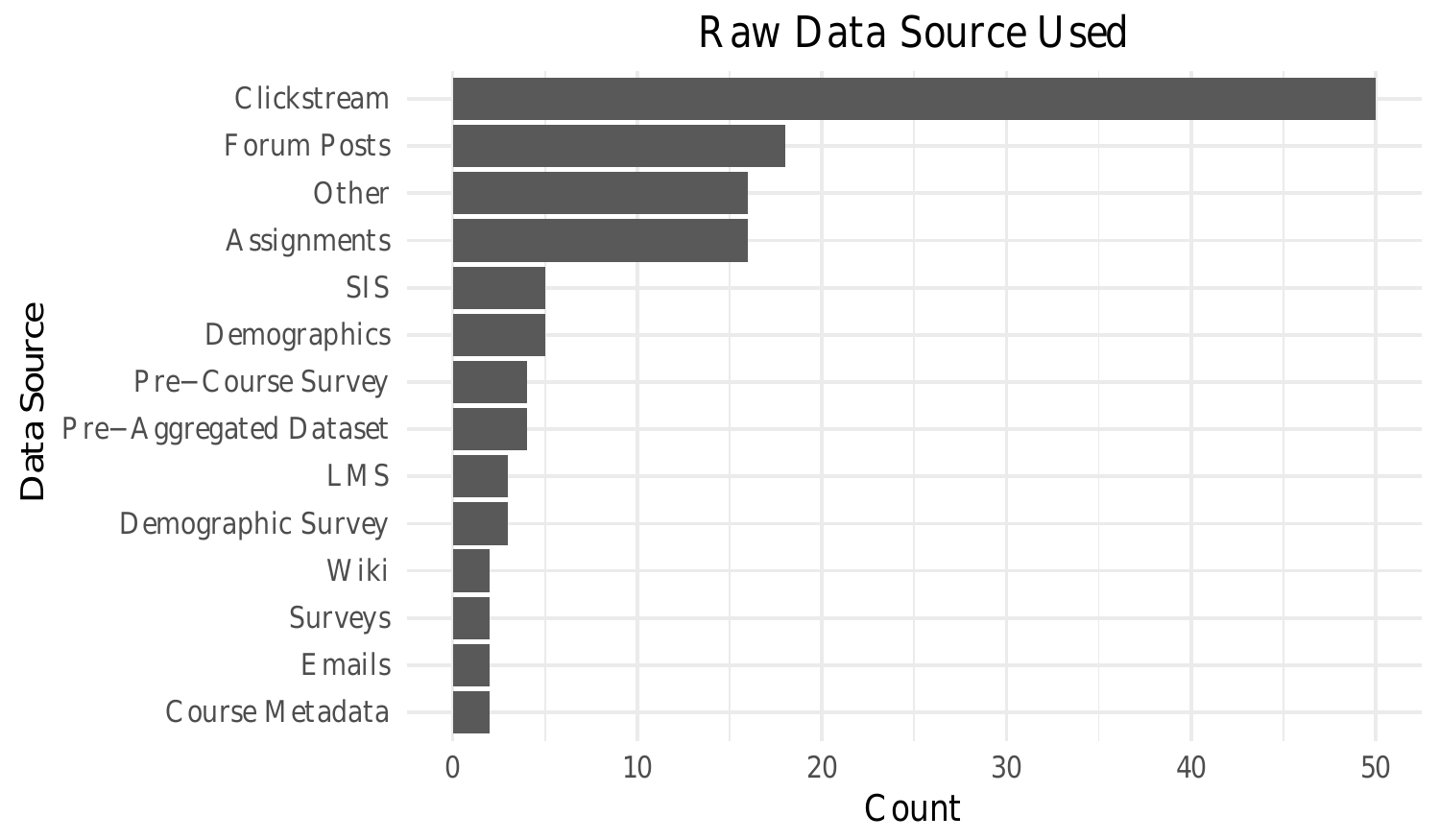}
    \caption{MOOC data sources used in predictive modeling research surveyed.}
    \label{fig:data_source}
\end{figure}

\begin{figure}
    \centering
    \includegraphics[width=0.8\textwidth]{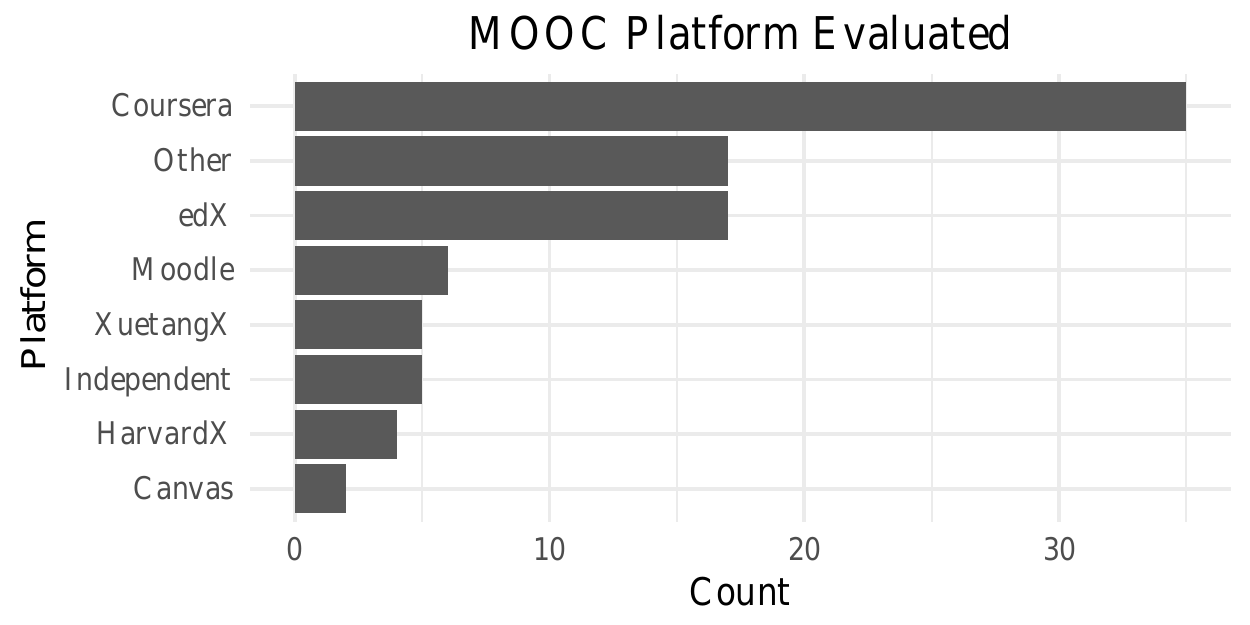}
    \caption{MOOC platforms evaluated in predictive modeling research surveyed. Research on various platforms largely reflects the  distribution of learners across these platforms. Note that certain ``platforms'' (e.g. HarvardX, XuentangX) may use software of vendor platforms (notably, edX) but do so in a way which is independent of that vendor platform.}
    \label{fig:platform}
\end{figure}

We provide data on the MOOC \textit{platforms} evaluated across work surveyed in Figure \ref{fig:platform}. These results reflect the dominance of the two largest MOOC providers, Coursera and edX \citep{Shah2018-cn}. Non-English MOOC platforms, such as the Chinese platform XuetangX, are less well represented in the work surveyed. As non-English platforms continue to grow, they should be researched more extensively: a substantial segment of the populations who stand to benefit most from global access to MOOCs are non-English speaking, and these learners are likely to differ from the population of English-speaking course takers.

Figure \ref{fig:data_source} demonstrates that, of the raw data sources discussed in Section \ref{sec:mooc-data}, clickstreams are the dominant raw data source for predictive modeling research in MOOCs. In one sense, this is unsurprising: clickstreams provide rich, granular data that the field is only beginning to harness the ability to represent in its full complexity. On the other hand, clickstreams are raw, semi-structured text files that require extensive human and computational effort to parse. Their formats are complex and sometimes inconsistent due to errors in platform server logging, and several levels of aggregation can be applied to a given entry (i.e., clickstream entries contain both session and user IDs, such that aggregating at both levels is not possible). In contrast, the other data formats shown in Figure \ref{fig:data_source} are typically provided as structured relational databases that can be access with simple SQL statements. The fact that clickstreams are so widely used, despite these barriers to accessing and utilizing this data, is a testament to their usefulness in predictive modeling. \cite{Gardner2018-ga} evaluates features generated from different data sources, comparing the predictiveness of clickstream features vs. forum- and assignment-based features; this work verifies that clickstream features are more effective predictors than forum- or assignment-based features when predicting dropout across the entire population of learners in a large sample of MOOCs.

We observe a growing ``long tail'' of additional data sources, which represents a continued trend toward combining other data sources with MOOC data to gather a more complete picture of learners. This is a useful development, but the privacy-protected nature of learner data often make it difficult to combine with other sources. Finally, we note that the forthcoming discussion in Section \ref{sec:meth-res-gaps} is relevant to the use of clickstream data. While clickstreams contain complex, potentially useful \textit{temporal} information about learner behavior over time, most modeling has been limited to simple counting-based representations of these temporal patterns \citep[with few exceptions; i.e. ][]{Fei2015-ea, Brooks2015-ej}. Much of the complexity contained in these interaction logs has likely not been captured with the research methods used to date.

\subsection{Feature Engineering Methods}

\subsubsection{Feature Types Used in Work Surveyed}

Feature engineering from these data sources is a focal point of MOOC research, and many advances in predictive modeling have hinged on clever or state-of-the-art feature extraction techniques, even when strikingly simple models are used. For example, \cite{Veeramachaneni2014-ug} combines a comprehensive set of crowd-sourced features with a simple penalized logistic regression model; subsequent work demonstrates that this model is capable of state-of-the-art prediction accuracy despite its algorithmic simplicity \cite{Taylor2014-hu}. As mentioned in our introduction to Section \ref{sec:lit-review}, there is a clear consensus that feature extraction is important to predictive modeling in MOOCs, and that future work should continue these investigations into feature engineering. Additionally, work which \textit{compares} the predictive usefulness of various feature sets in a rigorous, experimental way -- as in \cite{Crossley2016-ij} with activity- and NLP-based features, \cite{Brooks2015-ej} with demographics and activity features, and \cite{Gardner2018-ga} with activity, forum, and assignment features -- will be particularly useful as feature engineering continues to diversify. As Sinha et al. note, ``[t]he biggest limitation of most of these emerging works is that they focus solely on discussion forum behavior or video lecture activity, but do not fuse and take them into account'' \citep[][p.1]{Sinha2014-sr} -- the focus on using individual groups of features is holding back predictive modeling research in MOOCs.

Figure \ref{fig:model_categories} shows the broad categories used for taxonomizing the work surveyed here, which are largely (although not entirely) based on features. This data clearly demonstrates the dominance of activity-based feature extraction approaches. Two main factors explain this dominance: first, activity data is simply the most prevalent and fine-grained data available from MOOC platforms, and there are rich, complex patterns embedded in this data that the scientific community has correctly identified as important to explore. Second, activity-based \textit{outcomes} (i.e., dropout or stopout) have been a focus of MOOC research, as shown in Table \ref{tab:features-outcomes}. Activity features seem a necessary (if not sufficient) set of features for the task. As research begins to explore other outcomes beyond dropout and completion (such as learning), and as feature extraction becomes a less labor-intensive task perhaps due to open-sourced code or open MOOC data analysis frameworks \citep[e.g.][]{Gardner2018-hj}, it is likely that feature engineering will increasingly utilize other feature types either in addition to or instead of activity-based features.

\subsubsection{Input Features and Prediction Outcomes}~\label{sec:features-outcomes}

Because of the importance of feature engineering to the work of predictive modeling in MOOCs, and because this is the first large-scale survey of such work, we also provide detailed data on the relationship between model types and prediction outcomes used across work surveyed. The trends we observe suggest uneven exploration of different model types and student success outcomes across the work surveyed, suggesting both (a) a family of well-researched outcomes which we may be able to more reliably predict using insights from prior work, and (b) potential areas for further research.

\begin{figure}
    \centering
    \includegraphics[width = \textwidth]{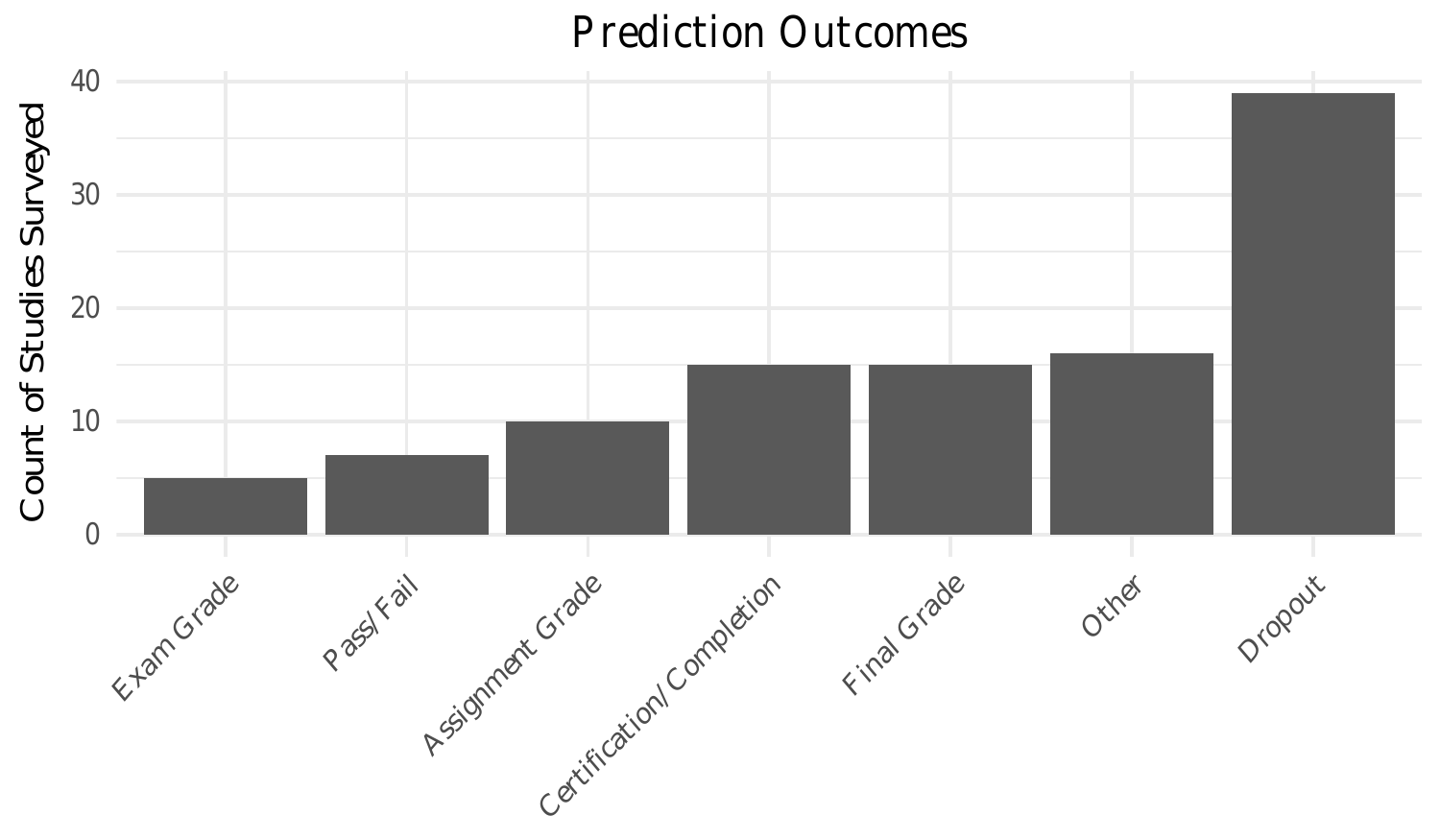}
    \caption{Student success outcomes predicted by works surveyed. When experiments predicted multiple outcomes, they were included in each category in this table, so the total across all groups exceeds the total number of works surveyed.}
    \label{fig:outcomes}
\end{figure}

Figure \ref{fig:outcomes} demonstrates that dropout prediction was more than twice as common as any other outcome predicted across our survey. 39 works attempted to predict some form of dropout or stopout. In contrast, outcomes related to completion, certification, grades, or other outcomes (e.g. level of engagement \citep{Cocea2007-eu}, ``healthy'' vs. ``unhealthy'' attrition \citep{Vitiello2017-fr}) were predicted less commonly and at similar frequencies. This largely reflects the current state of the MOOC landscape since 2012, discussed previously: concern about low completion rates prompted extensive research into the factors driving these rates.

We demonstrated in Figure \ref{fig:model_categories} that activity-based models were the most prevalent across our survey. Table \ref{tab:features-outcomes} adds further context to these groupings, demonstrating which model types were used to predict various student outcomes in MOOCs. This suggests more specific research gaps than those in Figure \ref{fig:outcomes}: for example, only one work surveyed used a cognitive modeling approach to predict completion \citep{Kizilcec2015-gu}, and only two used learning models to predict completion \citep{Jiang2014-vk, Qiu2016-ct}. Table \ref{tab:features-outcomes} demonstrates several such avenues for potential research in this and other areas as MOOC prediction moves beyond activity-based dropout modeling, the most common approach to date. 

\begin{table}[]
\centering
\begin{tabular}{p{0.1cm} p{1.2cm} p{1cm} p{0.8cm} p{0.8cm} p{1.2cm} p{1.2cm} p{0.7cm} p{0.6cm}}
\multicolumn{9}{c}{\textbf{Model Type}}  \\ \hline 
\multirow{7}{*}{\rotatebox[origin=c]{90}{\textbf{Outcomes}}} &  & \textbf{Activity} & \textbf{Text} & \textbf{Social} & \textbf{Cognitive} & \textbf{Learning} & \textbf{Dem.} & \textbf{Total} \\ \hline 
 & Academic & 15 & 7 & 6 & 7 & 15 & 11 & \textbf{61} \\
 & Completion & 9 & 5 & 3 & 1 & 2 & 6 & \textbf{26} \\
 & Dropout & 29 & 6 & 5 & 6 & 4 & 7 & \textbf{57} \\
 & Other & 11 & 3 & 1 & 5 & 5 & 1 & \textbf{26} \\
 & \textbf{Total} & \textbf{64} & \textbf{21} & \textbf{15} & \textbf{19} & \textbf{26} & \textbf{25} &
\end{tabular}
\caption{Model type (according to categories in Section \ref{sec:lit-review}) vs. prediction outcomes across works surveyed. When experiments considered a predictive model which could be considered multiply types, or predicted multiple outcomes, they were included in each category in this table, so cell totals exceed the total number of works surveyed. ``Academic'' outcomes includes: pass/fail, final grade, assignment grade, exam grade. ``Completion'' includes all metrics of course completion, e.g. certification, participation in final course module.}~\label{tab:features-outcomes}
\end{table}

Note that several outcomes in Figure \ref{fig:outcomes} are grouped into a single ``academic'' outcome category in Table \ref{tab:features-outcomes}. ``Pass/Fail'' is typically an indicator for whether a learner exceeded a predetermined final grade threshold for passing the course; ``Certification/Completion'' is typically an indicator for whether a learner officially completed all course requirements for an official certificate of completion (which sometimes, but not always, requires payment and identity verification).

\subsection{Modeling Algorithms}

\begin{figure}
    \centering
    \includegraphics[width=\textwidth]{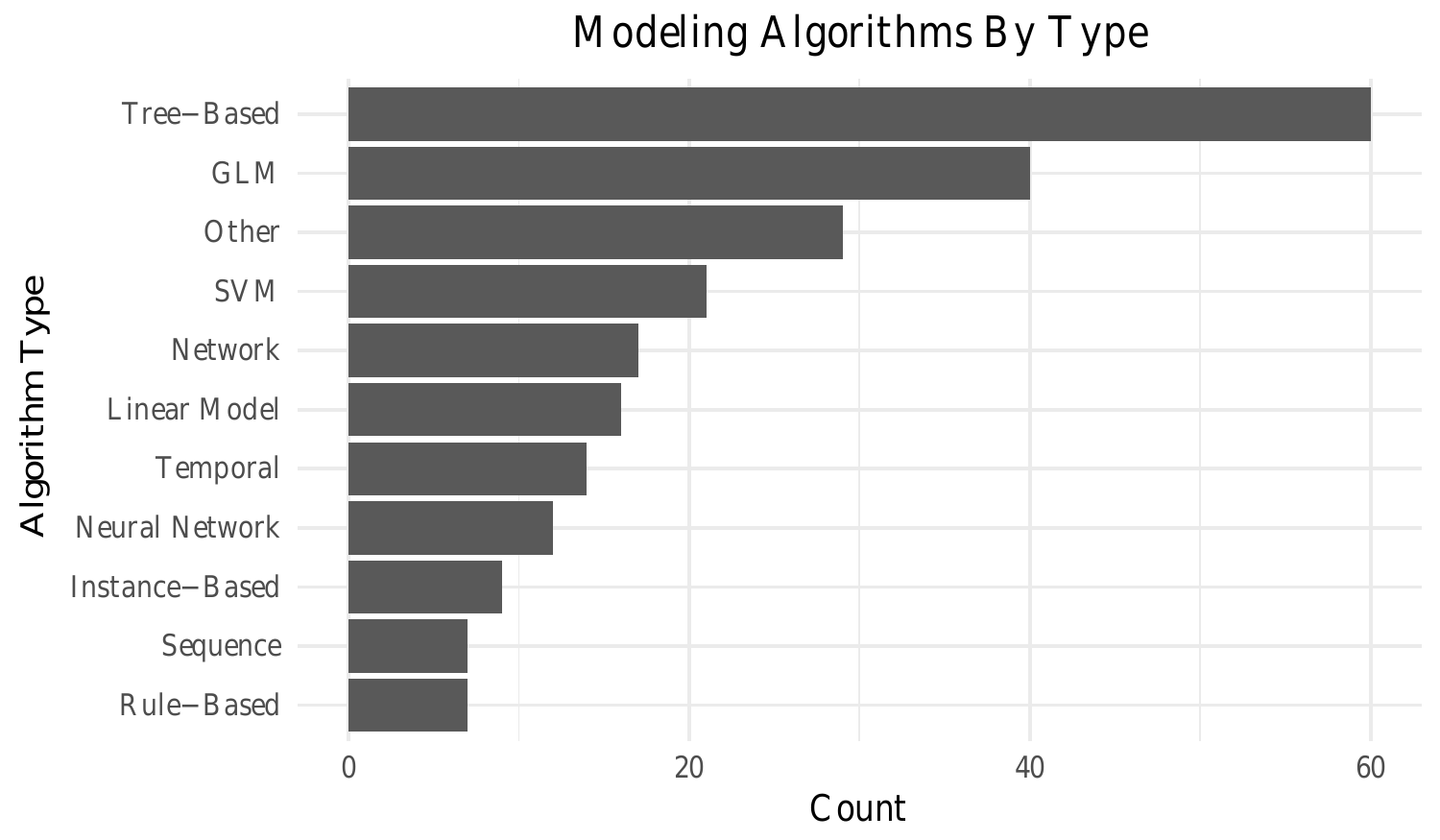}
    \caption{ABOVE: (a) Modeling algorithms used in works surveyed, by model type. Tree-based models appear to be particularly popular for their interpretability; Generalized Linear Models (GLMs) appear to be common because of their strong empirical performance and low bias. BELOW: (b) detailed breakdown of modeling algorithms used in works surveyed, by individual algorithm.}
    \includegraphics[width=\textwidth]{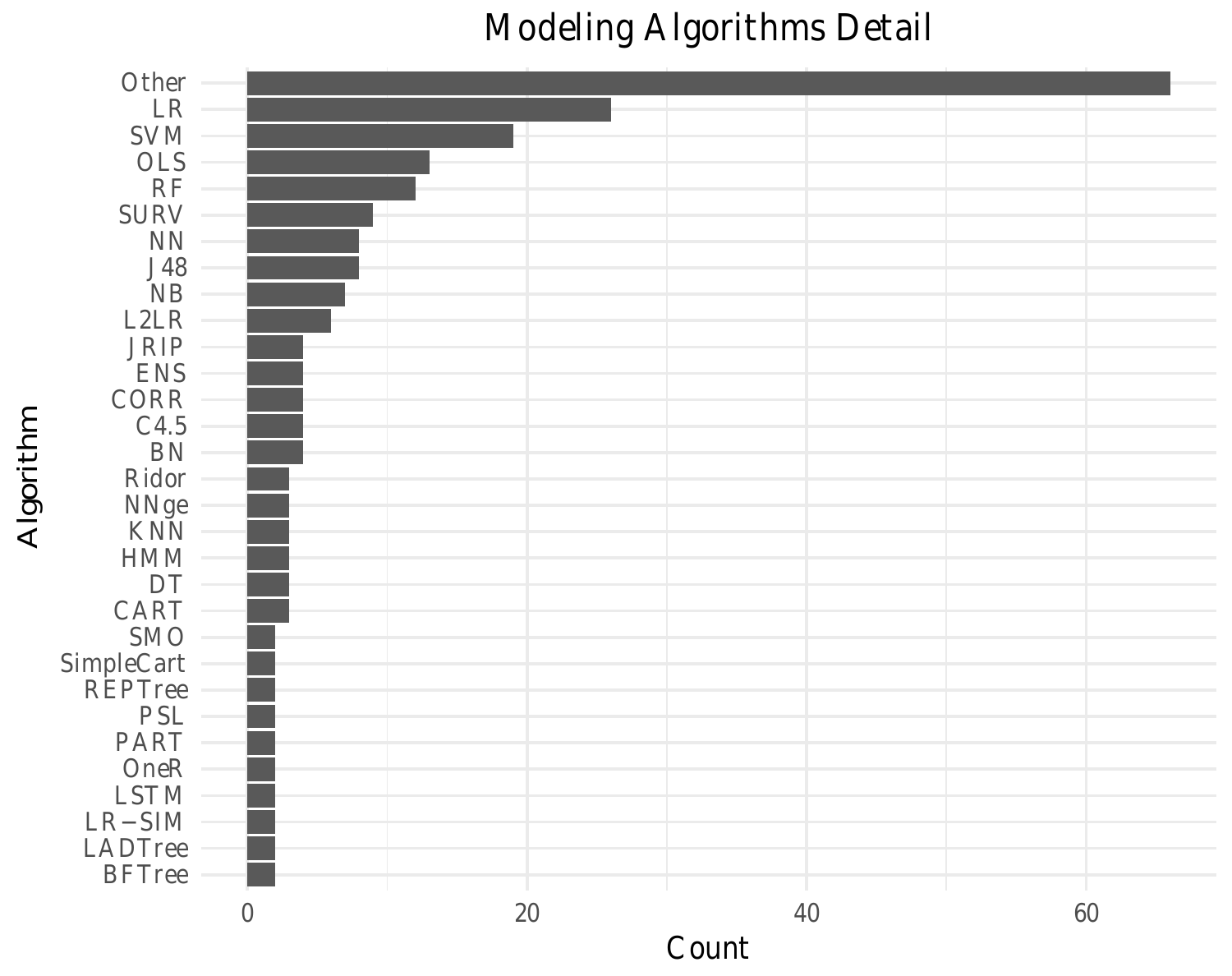}
    \label{fig:algorithms}
\end{figure}

The statistical models used to map features to predictions are a core component of predictive student modeling in MOOCs, but there is little prior synthesis of the findings of which algorithms are most widely used. Figure \ref{fig:algorithms} provides two perspectives on the modeling algorithms used across our survey.

First, the top panel, Figure \ref{fig:algorithms}a shows that tree-based models and generalized linear models are the most common techniques for predictive modeling in MOOCs. The prevalence of tree-based algorithms is due to several useful properties of these techniques: tree-based models can handle different data types (i.e., categorical, binary, and continuous) and are less susceptible to multicollinearity than linear models; they are relatively fast and simple to fit; they are nonparametric and make few assumptions about the underlying data while providing highly flexible models; and the results of these models are highly interpretable by visualization, inspection of decision rules, variable importance metrics etc.. Figure \ref{fig:algorithms}a shows that generalized linear models (GLMs) are also popular for MOOC learner modeling. This reflects several benefits of these models, in particular: GLMs empirically have achieved excellent performance across many large-scale MOOC modeling experiments; they are fast and simple to fit to data, requiring little or no hyperparameter tuning; and they produce interpretable output (which provides different information compared to tree-based models), including coefficients representing the magnitude and direction of association between each predictor and the response, and the statistical significance of these predictors.

Second, the lower panel, Figure \ref{fig:algorithms}b, shows the specific algorithms used across work surveyed, essentially disaggregating Figure \ref{fig:algorithms}a. Figure \ref{fig:algorithms}b shows how the dominance of tree-based algorithms largely obscures the lack of uniformity on which specific algorithms are used; of all tree-based algorithms considered, only random forests were used in more than 10 works surveyed. This makes it difficult to evaluate the effectiveness of any specific tree algorithm across our survey. In contrast, there are relatively few GLM algorithms adopted in the literature; logistic regression (LR) and L2-penalized logistic regression (``ridge'' regression, L2LR) account for almost all use of GLM algorithms. As noted above, GLMs, and L2LR in particular, generally achieve excellent performance when used with large and robust feature sets, despite their strong parametric assumptions about the underlying data.

Finally, Figure \ref{fig:algorithms} clearly reveals that there is a ``long tail'' of modeling techniques represented in the work surveyed here, with nearly half of the work surveyed using an algorithm which is not utilized in any other work (represented by `Other' in Figure \ref{fig:algorithms}a and \ref{fig:algorithms}b). In part, this represents an emphasis on novelty in published academic research; this is also indicative of a nascent field which has little consensus on the best approach to solving its prediction problems. We note that none of the algorithms in the work surveyed demonstrate performance which consistently exceeds all other algorithms, suggesting that there is indeed no single ``best'' algorithm \textit{a priori} for a given task or dataset \cite{Wolpert1997-wr}. Future work which compares and evaluates the fitness of various predictive modeling algorithms for different tasks in MOOC research would be appropriate at this stage; we advocate such work in Section \ref{sec:future-research} below. 

We observe that \textit{supervised} learning approaches dominate the literature, with few examples of unsupervised approaches; this is likely due to the fact that many of the outcomes (i.e., dropout, certification, pass/fail, grades) are observable for all learners, making unsupervised techniques unnecessary for many  of the prediction tasks addressed by research to date.

\subsection{Model Evaluation Metrics}

\begin{figure}
    \centering
    \includegraphics[width=\textwidth]{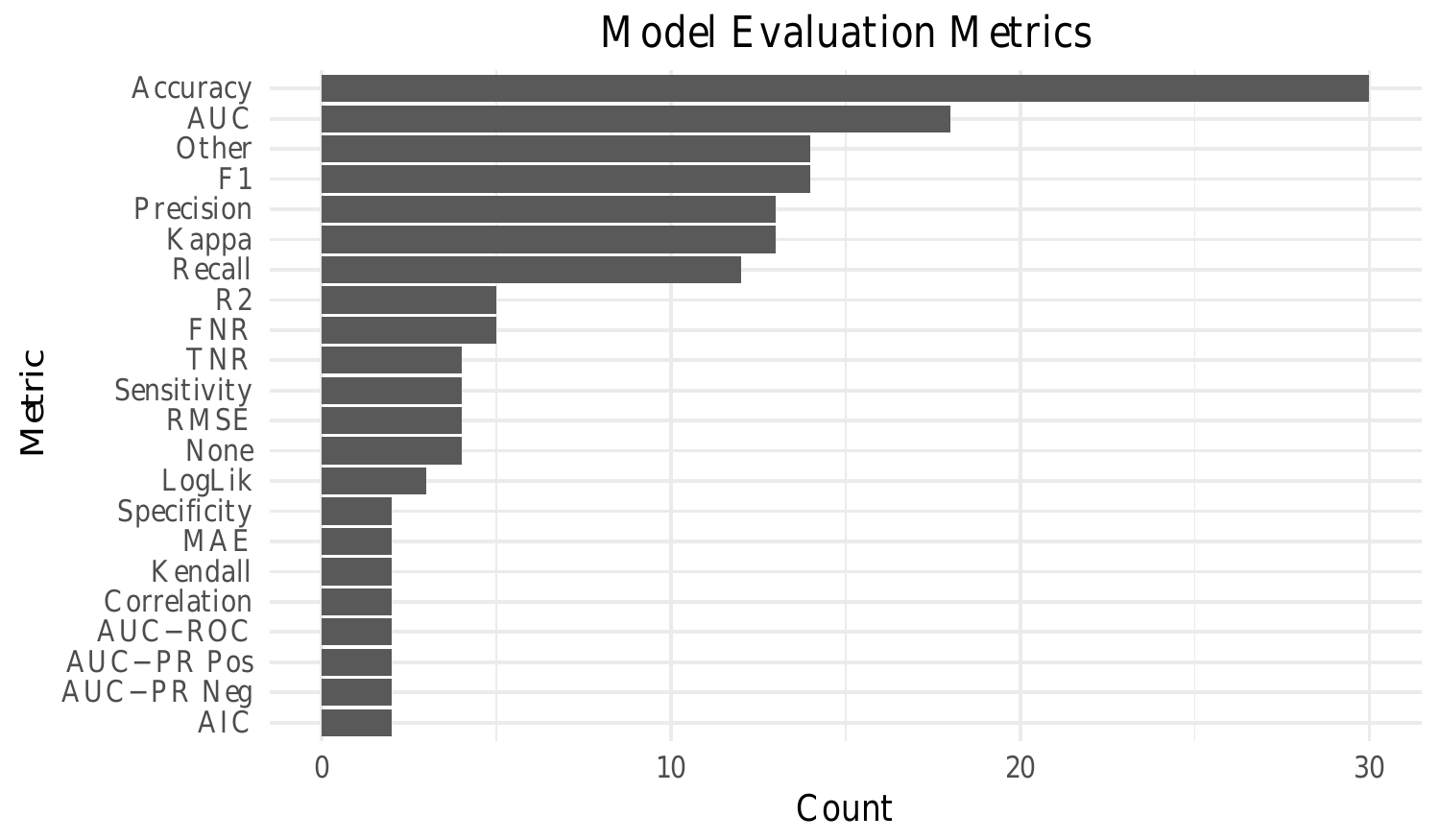}
    \caption{Evaluation metrics reported for predictive modeling experiments in work surveyed. Note that individual works are counted multiple times when results are presented according to multiple model performance metrics. Selected Abbreviations: FNR/TNR = False/True Negative Rate; RMSE = root mean squared error; LogLik = Log-Likelihood; R2 = r-squared.}
    \label{fig:evaluation_metrics}
\end{figure}

Our data also reveal a considerable lack of agreement about which model evaluation \textit{metrics} to use in MOOCs, shown in Figure \ref{fig:evaluation_metrics}. Compared to the analysis of algorithms and data sources above, this data reveals a slightly stronger consensus around a smaller set of evaluation metrics, most notably accuracy (ACC), Area Under the Receiver Operating Characteristic Curve (AUC), precision (also called positive predictive value) (PREC), and recall (REC) (also called true positive rate, sensitivity, or probability of detection). Strictly speaking, a diversity of metrics is not a problem -- different metrics measure different aspects of predictive quality, which vary depending on the task and research goals -- but this lack of a consistent baseline leaves readers unable to compare performance across otherwise-similar studies which report different performance metrics. Reporting \textit{several} metrics would often give a more complete picture of model performance and allow for easier comparison across studies, while still allowing researchers to examine performance according to their metric(s) of interest. Open data or open replication frameworks would allow for more nuanced comparison and would shift the burden from purely on the researcher, to allowing reviewers and critical readers to inspect results using any performance metric of interest.

10 of the works surveyed -- over 10\% -- report classification accuracy as the \textit{only} model performance metric. We consider this practice particularly concerning. Accuracy is useful and interpretable for many readers, but it can be a misleading measurement of prediction quality with highly imbalanced outcome classes. This scenario is very common in MOOCs (i.e., most students drop out, do not certify, etc.). Accuracy is also threshold-dependent, while other metrics, such as Area Under the Receiver Operating Characteristic, measure performance over all possible thresholds. While accuracy is useful as an interpretable metric for readers, it is often difficult to assess the value of work which only reports performance using accuracy. The practice of only reporting accuracy should be discouraged, as computing additional performance metrics from the data used to compute accuracy (namely, predicted labels and class labels) requires minimal additional effort (sensitivity, specificity, F1, Fleiss' Kappa, and several other metrics can be computed from these labels).

Additionally, it is important to note that the appropriate model evaluation metric often depends on both the outcome being measured and on the unique goals of a predictive modeling experiment. For example, in a dropout modeling experiment where the goal is to provide an inexpensive, simple intervention to learners (such as a reminder or encouragement), recall might be an appropriate model evaluation metric; in contrast, when the goal is to provide an expensive or resource-intensive support to predicted dropouts, precision might be a better choice.

Together, the data source, feature extraction method, statistical modeling algorithm, and evaluation metric reflect the \textit{accuracy} dimension of predictive student models introduced in Figure \ref{fig:modeling-dimensions}. A key element of any usage of predictive student models requires that these models are effective at predicting the outcome of interest; research into -- and methodological progress in -- each of these areas (feature extraction, modeling, and evaluation) stand to substantially improve the accuracy of future predictive MOOC models.

\section{Methodological and Research Gaps}\label{sec:meth-res-gaps}

In this section, we critically review the existing research body on predictive models of student success in MOOCs. In particular, we highlight (a) areas where the methods of prior research or the interpretation of its results are biased toward specific populations or overly prone to statistical errors; and (b) opportunities for future research toward modeling and understanding learner behavior in MOOCs. Because these issues are often two sides of the same coin (methodological gaps imply future research opportunities), we discuss them together. Additional opportunities for future research are discussed in Section \ref{sec:future-research}.

\subsection{``Small'' Data and Experimental Population Filtering}\label{sec:subpopulations}

Many of the challenges discussed in this section point to the difficulty of comparing findings across experiments. Because many MOOCs are at least superficially similar to each other and offered in similar contexts, this type of comparison is theoretically possible: making these comparisons would be neither unreasonable nor difficult, and would allow evidence for or against specific predictive modeling techniques to accumulate across experiments. However, under existing research methods, experimental populations are often not comparable. 

The use of small, highly-subsetted experimental populations in prior work is one way in which its generalizability is limited. Often, predictive experiments in MOOCs identify a subpopulation of learners on which the analysis is conducted. Unfortunately, these subpopulations are often so divergent that the results from one experimental subpopulation to another could be entirely different. Examples of experimental subpopulations from work surveyed are shown in Table \ref{tab:subpopulations}. We note that 40 of the works surveyed, or 46\%, filtered the sample from the total available population of registrants or participants in some way.

\begin{table}[]
\centering
\label{tab:subpopulations}
\begin{tabular}{p{2.75cm} p{6.25cm} p{2cm}}
 \hline Study & Subpopulation & \% of Enrollees \\ \hline
\cite{Perna2014-cp} & Registered after official course start date and no more than two months after course end date & 90\% \\
\cite{Halawa2014-pj} & Joined in the first 10 days of the course and have viewed at least one video & Not Reported \\
\cite{Fei2015-ea} & Students with at least one interaction as measured by 7 features used & 46.7\% \\
\cite{He2015-ab} & Students who submitted at least 1 assignment each week. & 13\% \\
\cite{Greene2015-no} & Completed pre-course survey and completed the first end-of-unit exam. & 11.4\% \\
\cite{Yang2013-zq} & Posted at least once in discussion forum by seventh course week. & 6.3\% \\ 
\cite{Robinson2016-yr} & Started in first two weeks; completed pre-course survey; saw utility value of course; fluent in writing English; intends to complete course; and wrote more than one word on survey. & \textless 5\% \\ \hline
\end{tabular}
\caption{A sample of experimental subpopulations in works surveyed. Percentage of total enrolled students is shown. Such divergent filtering criteria and small, nonoverlapping subpopulations make comparing the results of different predictive work difficult.}
\end{table}

Several comments are warranted here. 

First, there is tremendous diversity in the subpopulations evaluated by different works, and it is difficult, if not impossible, to compare findings of otherwise-similar experiments across studies. We simply do not know, for example, how the students who submitted at least one weekly assignment in \cite{He2015-ab} might compare to students who watched a video and submitted a problem in \cite{Li2016-gx} or to those who completed the survey with the relevant characteristics considered by \cite{Robinson2016-yr}. 

Second, filtering the population so significantly -- as Table \ref{tab:subpopulations} shows, many experiments which share this data reveal that over 80\% of MOOC participants are excluded from their analysis -- moves research further from the goal of understanding large segments of the learner population. While it is useful, for example, to know how natural language features and unigram frequencies in \cite{Robinson2016-yr} can be used to predict persistence, this data is of little practical use if it only applies to fluent English-speakers who see the value of the course, intend to complete it, and completed a pre-course survey with more than one word. Indeed, we would expect such learners to be quite different from the average or overall population in such a course, and such work gives us little information about the broader population.\footnote{This concern is similar to that raised in \cite{Henrich2010-hs} in the context of psychological research; as Henrich et al. argue, such sampling bias could have true and significant consequences for the generalizability of these findings.} Previous educational data mining research suggests that affect detection models, for example, do \textit{not} transfer across even regional and demographic boundaries within the United States \citep{Ocumpaugh2014-cf}. It seems even less likely that models trained on subpopulations of globally diverse MOOC learners, for example natural language models which only evaluate courses conducted in English, would generalize effectively across behavioral subgroups. At the very least, presenting the results both in terms of a small subpopulation \textit{and} in terms of the entire course population would provide a useful point of reference.

Third, these highly-subsetted experimental populations are often themselves a ``sample of samples'', evaluating just one or a few MOOCs which may or may not be representative of the larger population of MOOCs. Conducting research using these types of populations make it difficult to determine how such work might generalize even to the same subpopulation in other courses. Table \ref{tab:num_courses} shows that over 50\% of the works surveyed evaluated just one MOOC, with fewer than 20\% of these works evaluating 10 or more courses. At best, highly-subsetted populations of an already narrow sample (of the overall MOOC population) can be taken as promising avenues for future research; it would be a mistake to consider the findings of such research fully resolved conclusions. We should be particularly concerned about works which publish ``statistically significant'' results for single-course populations in a small and highly-contingent subpopulation: these analyses can be subject to high ``researcher degrees of freedom'' \citep{Gelman2013-ok}, and the extent to which these degrees of freedom were exploited during data analysis is rarely reported in published research. This bias may be compounded when predictive models are evaluated on the same course that is used to fit them \citep{Boyer2015-lo, Whitehill2017-tt}.

\begin{table}[]
\centering
\begin{tabular}{ll}
\hline
Number of Courses & Count \\ \hline
1-5 & 63 \\
6-10 & 4 \\
11-15 & 6 \\
16-20 & 1 \\
21-30 & 1 \\
31-40 & 5 \\
40-50 & 1 \\
51-100 & 0 \\
100-150 & 1 \\ \hline
\end{tabular}
\caption{Number of MOOCs evaluated across research surveyed. Most studies (70\% of work surveyed) evaluate data from 1-5 courses.}~\label{tab:num_courses}
\end{table}

Having discussed the limitations and challenges raised by work which utilizes such specific experimental populations, we recognize, of course, the value of such research, even with its limits, and the reasons for doing so in practice. Our intent is not to single out the authors of any one particular study; indeed, the fact that this applies so broadly to many of the most highly-cited works in the field suggests that even many of the most substantial contributors to the field have conducted such research. Many of these works reflect early exploratory research into MOOCs, and were initial efforts at understanding \textit{any} cross-section of this novel population.  We simply note that these limits are often not acknowledged by the broader research community when interpreting these results, and that further research is needed to explore the generalizability of these findings and ensure that the field's knowledge base is constructed on firm ground as the field grows and matures.

\subsection{Model Evaluation, Comparison, and Replication}\label{sec:mod-eval}

A second area where substantial research gaps exist is in the evaluation, comparison, and replication of the predictive models of student success in MOOCs. As work on predictive modeling has expanded across all domains, a substantial research base has emerged on techniques for comparing and evaluating the results of predictive modeling experiments. We find that the work surveyed often lags behind these accepted standards and methods for practice, which can be applied to predictive models in any domain. This also raises concerns about the \textit{accuracy} dimension of these models, particularly when applied to unseen data.

\subsubsection{Multiple Comparisons and Statistical Testing}

There is concern in the broader statistical community about issues of multiple comparisons in model evaluation, particularly when applied to large spaces of potential statistical models. The field has begun to move beyond these concerns through to the adoption of simple (if conservative) tehniques for accounting for the many comparisons performed over the course of an experiment (i.e., the methods of Bonferonni or \cite{Benjamini1995-nu}, or techniques specific to the evaluation of machine learning models outlined in \cite{Demsar2006-cx}). Bayesian methods have also been increasingly adopted for inference and data analysis, in part due to their robustness in cases of multiple comparisons \cite{Benavoli2017-ff, Gelman2012-ys}. 

However, almost none of the work surveyed utilized appropriate significance testing techniques (according to the standards of \cite{Demsar2006-cx} or \cite{Benavoli2017-ff}). \cite{Molina2012-xj} was the only exception, based on our reading of these works, but Molina evaluates traditional courses managed in Moodle, not MOOCs. We also found no acknowledgement of concerns about multiple comparisons in interpreting the statistical significance of results in any work. This lack of concern exists in spite of the fact that 18 of the works surveyed (20\%) reported evaluating more than 20 models as shown in Table \ref{tab:num_models}, which means that at least one Type I Error would be \textit{expected} for a single test at a 5\% significance level using a traditional hypothesis test to compare models. It is possible that works which reported fewer than 20 models evaluated additional predictive models in the course of their experiments, exposing these experiments to an inflated risk as well.

\begin{table}[]
\centering
\begin{tabular}{ll}
\hline 
Number of Models & Count \\ \hline 
1-5 & 34 \\
6-10 & 9 \\
11-20 & 5 \\
21-30 & 3 \\
31-40 & 2 \\
41-50 & 2 \\
51-100 & 4 \\
101-500 & 4 \\
501-1000 & 1 \\
\textgreater 1000 & 2 \\ \hline 
\end{tabular}

\caption{Counts of the number of predictive models reported to have been fitted/compared within each of the studies evaluated. 42 experiments -- 48\% of work surveyed -- reported evaluating more than 20 different predictive models, raising clear methodological concerns about multiple comparisons.}~\label{tab:num_models}
\end{table}

There are clear reasons why these methodological concerns emerged in the first place. First, a complete lack of testing fails to quantifiably evaluate the findings of a predictive model. Some form of evaluation is needed to quantify the degree to which we might attribute observed differences in performance to chance versus to a ``better'' model or feature extraction technique. This is especially important given the small samples of courses in works surveyed shown in Table \ref{tab:num_courses}. Second, even when statistical testing is used, often these tests require specific corrections when applied for predictive model evaluation. Many common statistical tests, such as the Student's $t$-test or Analysis of Variance (ANOVA), are not appropriate or calibrated for testing predictive models \cite{Demsar2006-cx, Dietterich1998-vh}. It is possible that the concerns which motivated these approaches may have been realized in many of the works surveyed\footnote{Some corrections, such as the Bonferonni correction, can be applied by readers directly by simply multiplying the reported $p$-value by the number of comparisons; however, even this depends on the researcher self-reporting the number of models considered. It is unlikely that the total number of models considered over the scope of an entire experiment are reported in most published research.}. Even the large number of models \textit{reported} in the work surveyed (note that at least $\frac{1}{3}$ of the works did not report the total number of models evaluated) suggest that inferential errors caused by uncorrected multiple comparisons may lurk in the current knowledge base of student success models. This lack of replicability of most work (discussed below), combined with the ``file drawer problem'' wherein null results are rarely published \cite{Rosenthal1979-nr}, make it particularly difficult to determine when these Type I errors may have occurred or how prevalent they may truly be in the field of predictive MOOC modeling.

While the appropriate technique(s) for model evaluation vary based on the nature of the comparison (i.e., two models vs. many models; a single dataset vs. many datasets), these procedures do exist and are often simply ignored in predictive modeling research in MOOCs. These procedures are discussed in detail in a future work regarding predictive model evaluation in MOOCs; we refer the reader to that work or to \citep{Benavoli2017-ff, Dietterich1998-vh, Demsar2006-cx} for further details. 

\subsubsection{Cross-Validation for Model Inference}

Particularly relevant in this discussion are the specific limitations of using the results of cross-validation to compare and draw inferences about the performance of predictive models. Average cross-validation performance was used to evaluate and compare the performance of 31 studies, or nearly 40\% of the work surveyed. Again, problems with this procedure have been well-studied. Utilizing average cross-validated performance with unadjusted statistical tests (such as a paired $t$-test) makes such research susceptible to both high Type I error rates (higher than expected probability of concluding that there is a significant difference in performance when none exists) and low power (low ability to discern true differences in performance when they do exist) \citep{Dietterich1998-vh, Bouckaert2004-oy}. These issues are discussed specifically in the context of model evaluation in MOOCs in \cite{Gardner2018-ga}. Works which evaluate many predictive models -- effectively conducting large numbers of multiple comparisons -- inflate this risk. This exposes predictive modeling research in MOOCs to serious and preventable concerns about the reproducibility of its findings, at a critical time when the field's work is growing in both visibility and practical significance.

\subsubsection{Replication of Predictive Modeling Experiments}

Taken in sum, the two challenges outlined above -- multiple comparisons and a lack of rigor in model evaluation -- point to a third challenge in predictive research in MOOCs: replication. We note with some concern that there is a dearth of replication research in MOOCs in particular, and in the field of education in general \citep{Makel2014-kd}. This means that despite the concerns, outlined above, about the inferential and sampling procedures often used in MOOC research, we are unable to estimate the impact of these procedures on the generalizability of many published findings. Of the work surveyed for this evaluation, none was a replication of prior work by new authors, although in limited cases (a) original authors reproduced their analyses on new MOOC datasets \citep[e.g.][]{Rose2014-mk}, or (b) new authors attempted to at least compare their work to algorithms used in others' work as a baseline  \citep[e.g.][]{Fei2015-ea}. An initial attempt at replication in MOOC models is shown in \cite{Andres2016-ju, Andres2018-mt}, but these works replicate predictive models as relatively limited production-rule analyses and do not replicate the predictive models themselves (i.e., by controlling for covariates). Exact replication should be relatively more tractable in MOOC research than in other fields: MOOC data is largely consistent within (and even across) the two major platforms, Coursera and edX. The largest apparent barriers to replication are (a) lack of access to data; (b) lack of clarity in published descriptions of experimental methods; and (c) the lack of incentive to replicate previously-published research. Extensive work on the challenges of reproducible computational research \citep{Peng2011-az}, best practices for conducting computational research \citep{Stodden2013-bh}, and tools or software for facilitating such research \citep{Kitzes2017-pf} provide a foundation for future efforts in the field.

Particularly with the rapid proliferation of different approaches to predictive modeling in MOOCs, replication would provide a useful basis for comparing these approaches. Work that exactly implements the methods from another experiment has been called \textit{direct replication} by \cite{Donoho2015-aq}; such research is extremely uncommon in educational research \cite{Makel2014-kd} but is needed. For example, while multiple existing studies might use an SVM and compare these results, the comparisons often ignore important differences in hyperparameter tuning, kernel selection, regularization, and feature selection which can have genuine effects on the performance of these algorithms across experiments. Future work which evaluates predictive models using \textit{multiple} outcome metrics would also give a more complete picture of their performance  \citep[i.e.][]{Fei2015-ea}, even when authors cannot or choose not to openly share their code or data for replication. We note that \cite{Kitzes2017-pf} provides several useful case studies for addressing computational reproducibility across several domains, including domains which require working with privacy-restricted data (e.g. health care, nuclear physics); recent work in MOOCs has also made progress toward sociotechnical solutions to this problem \citep{Gardner2018-hj}.

\subsubsection{Toward the ``State of the Art''}

In conclusion, the generalizability of many results in the work surveyed is seriously called into question. Future work which (a) adopts more effective inference and evaluation techniques which are robust to multiple comparisons and are appropriate for the evaluation of predictive models and (b) replicates the findings of prior work on new, larger data would make a valuable contribution to the pursuit of robust and generalizable knowledge about predictive modeling in MOOCs. 

We note that a particular consequence of the analysis of the current section and Section \ref{sec:subpopulations} is that they point to a confluence of factors which make it difficult, if not impossible, to reliably identify the ``state of the art'' in predictive models of student success in MOOCs. This extends to evaluations of both the best feature engineering methods and of statistical modeling techniques. Because of the large differences between the subpopulations evaluated, the model evaluation metrics, and the statistical methods for evaluating experimental results (if any), comparisons across studies to determine which methods are most effective are tenuous at best. We can make observations about the popularity of various techniques, and can note based on the current survey that activity-based features are the most commonly used, followed by text-based features. However, in order to draw reliable conclusions about which methods are truly the ``state of the art'' in student performance prediction, we would need one or more large, highly-representative, shared benchmarking datasets (and, ideally, infrastructure or tools for executing, sharing, and replicating experiments run on this dataset). As noted previously, the MOOC Replication Framework (MORF)\footnote{\url{educational-technology-collective.github.io/morf/}} and DataStage\footnote{\url{https://datastage.stanford.edu/}} represent possible solutions to conduct such comparisons in future work to truly determine the state of the art in MOOC student performance prediction.

\subsection{Realistic Experimental Contexts}\label{sec:realistic-experiments}

A second area in which methodology and experimentation in MOOC modeling stands to grow is the context in which predictive modeling experiments are conducted. In particular, we advocate the use of realistic experimental contexts in future work; the state of the practice largely produces models which are not \textit{actionable}. 

Building actionable predictive models to support downstream support and intervention is the stated aim of much of the work surveyed -- these works often explicitly describe the aspirational use of their predictive models as the linchpins of ``early warning'' systems for ``at-risk'' students. Some works surveyed describe planned or hypothetical interventions based on such models; one utilized a student-initiated micro-commitment intervention and explored using student commitment as a predictor of assignment submission \cite{Cheng2013-xe}, and a single work surveyed actually utilized predictive models for live adaptive interventions \cite{Whitehill2015-ap}. 

However, much of the work surveyed is simply not possible to implement in an active course -- we call such experimental contexts not \textit{realistic}. A realistic context matches the situation in which predictive models would be employed for active use in MOOCs, particularly with respect to the information available at the time of prediction. Many utilize \textit{post hoc} prediction architectures, where (a) model-fitting requires labels which are not knowable until a course completes, and (b) model evaluation takes place by evaluating test predictions made on the same course used for training -- not a disjoint future course. These contexts do not match those in which a real-time predictive model would be used: for example, dropout labels are not known at the time of training and prediction if a course is still in progress; by definition, a users' dropout status is not knowable until the course completes. Of the works surveyed, only \cite{Ashenafi2016-df, Bote-Lorenzo2017-yh, Boyer2015-lo, Brooks2015-ej,  Brooks2015-fh, He2015-ab, Kizilcec2015-gu, Wen2014-xr, Whitehill2015-ap, Whitehill2017-tt} -- fewer than 10\% -- examine prediction architectures in which the test predictions could be made for an incomplete course (either by training and predicting on different iterations/courses and using transfer learning, or using some form of proxy labeling).

The degree to which the prediction context, particularly same-course evaluation vs. future-course evaluation, may bias results is unclear. \cite{Veeramachaneni2014-ug} find that predicting on future courses generally achieves lower performance than same-course prediction, and that second-to-third transfer is more accurate than first-to-third. \cite{Whitehill2017-tt} examine a variety of prediction architectures and find that while same-course (\textit{post hoc}) prediction architectures optimistically bias estimates of model performance, \textit{in situ} proxy labeling achieves comparable performance. \cite{He2015-ab} find that ``prediction models trained on a first offering work well on a second offering'', with such models achieving an AUC of 0.8 using only one week of data when predicting on a future iteration. \cite{Evans2016-gj} shows that users engage with later runs differently from the way they engage with earlier runs in an analysis of 44 MOOCs, suggesting that such transfer would be less effective than with a model trained on another non-first run (i.e., training on second iteration, predicting on third iteration). Model transfer to future courses is also evaluated in \cite{Brooks2015-ej}, which achieves an AUC of 0.9 while predicting on future runs of a MOOC. These works collectively suggest that same-course training and prediction may optimistically bias results, but that accurate prediction on future iterations is possible and that multiple methods for such prediction exist.

Of course, real-time intervention is not the goal of \textit{all} predictive modeling research in MOOCs. In the case of many explanatory/inferential works, the goal of model-fitting is simply data understanding. In such cases, the issues highlighted above are less relevant. However, for any tasks which do indeed require real-time model-fitting or prediction -- which appears to be the ``gold standard'' for predictive research in MOOCs and the ultimate goal of many of the works surveyed -- utilizing techniques which are adaptable to such contexts is a necessity. Without using these architectures, we are left wondering whether the predictive performance achieved by many otherwise-promising works could be achieved under the constraints of real-time prediction or model-fitting.

Our goal in this section is not to suggest that prior work is useless, or even incorrect -- we believe that the search for effective predictive modeling techniques is an iterative process that requires initial experimentation and exploration, even in laboratory contexts which do not fully mimic real-world constraints -- but it certainly suggests a promising avenue for future research, which might be able to test previous feature engineering and modeling approaches, re-architected in ways that allow for model training and prediction in ``live'' environments.

An additional methodology that appears particularly useful for efficient real-world model training are incremental or pre-training approaches, which can efficiently update predictive models to incorporate new data without requiring additional passes over previously-seen data. Such techniques have been demonstrated in the incremental training utilized in \cite{Kotsiantis2010-hs, Sanchez-Santillan2016-ml}; and in the pre-training techniques used to incrementally grow the neural network models in \cite{Whitehill2017-tt}. We hope that future research adopts the use of incremental techniques, which would allow for smoother adoption of predictive models in practice.

Just as Section \ref{sec:subpopulations} highlighted how a large portion of the work surveyed fails to consider or model massive segments of the learner population, this section indicates how much of this research may fail to provide \textit{accurate} or \textit{actionable} insights for the segments it does evaluate.

\section{Opportunities for Future Research}\label{sec:future-research}

Several of the trends and methodological gaps outlined above directly suggest areas for future research. As we note above, this includes work which examines large, unfiltered, and multi-MOOC experimental populations; work applying rigorous model evaluation and comparison tests to identify effective feature engineering techniques and algorithms for prediction (including, especially, when such approaches may be statistically indistinguishable in terms of their predictive performance); and work utilizing training and prediction contexts which match those in which a predictive model might be deployed in a live course environment.

Our survey identifies four research gaps in addition to those described above: (1) adoption of temporal modeling techniques, (2) bridging the ``two cultures'' of statistical modeling in MOOC research, (3) \textit{theory-building} MOOC research, and (4) modeling long-term student success in MOOCs.

\subsection{Utilizing Temporal Modeling}\label{sec:temporal-modeling}

There is a clear temporal element to prediction in MOOCs: many courses are offered using a cohort-based model (for example, with new cohorts beginning at monthly intervals); course activity and learning takes place over time, with most courses lasting several weeks; data is collected incrementally, with little usage data being available during the early phases of a course and more data collected as it progresses; learner behavior evolves over the duration of a MOOC. This suggests that models which can account for and explicitly model the complex, time-dependent patterns in MOOC learner data are likely to form a more complete picture of this behavior than those which ignore the element of time. However, research to date has been limited in its use of temporal modeling techniques.

Most prior research which does account for temporality falls into two broad groups. (1) One group utilizes ``weekly'' feature sets to broadly capture separate collections of features over time periods, typically for each week of a course. Many of the works surveyed here utilize this approach, e.g. \cite{Kloft2014-kb, Vitiello2017-fr}.  \cite{Xing2016-le} refer to this as ``appended'' feature extraction. While this type of modeling does capture different features over time, it does not explicitly model these features as being captured sequentially, and treats those predictors as otherwise independent when they are actually related across time steps \citep{Wang2016-yc}. (2) A second broad class of work utilizes survival models. This includes \cite{Rose2014-mk, Wen2014-xr, Yang2013-zq, Yang2015-gy, Yang2014-ka}. Many of the methods used in these experiments, as the Cox Proportional Hazards Model for survival analysis and logistic regression, are forced to make the statistical assumption that student dropout probability at different time steps is independent \cite{Wang2016-yc} -- an assumption which is almost certainly violated, and which limits these models' ability to model correlation between student dropout probabilities at different steps over time.

Attempts to capture more complex temporal patterns in MOOC data have been limited. Hidden Markov Modeling has been used for some dropout models, most notably in \cite{Balakrishnan2013-ty}, but this is a generalized form of sequence modeling, not strictly a time-series methodology. A nonlinear state space model is used to capture longer-term information in student interaction sequences for dropout prediction in \cite{Wang2016-yc}. Some work has explored the use of higher-order time series data, utilizing n-gram models of feature sets or behavioral patterns \cite{Brooks2015-ej, Li2017-yd}. \cite{Fei2015-ea} explore the use of a form of complex neural network model, a Long Short-Term Memory (LSTM) Network. This LSTM model takes as inputs sequences of weekly feature vectors, and is used to predict dropout in this context. 

Future work which explores these approaches more deeply (such as by exploring other survival modeling approaches such as random survival forests \cite{Ishwaran2008-wb}), or which applies other time series approaches, would be valuable and is likely to uncover both informative patterns in data, and gains in predictive modeling performance, improving both the accuracy and theory-building components of future models.

\subsection{Bridging the ``Two Cultures''}\label{sec:two-cultures}

Another significant opportunity for future MOOC research is work which unites highly complex, predictive models with techniques for understanding and inspecting the relationships these models uncover, increasing the theory-buildingness and actionability of these models without sacrificing accuracy.

In his seminal 2001 essay \textit{Statistical Modeling: The Two Cultures}, Leo Breiman argued that the field of statistics was (at the time) divided between a \textit{data modeling culture}, concerned primarily with understanding the underlying data generation processes and which emphasized the use of inspectable, generative models such as linear regression; and an \textit{algorithmic culture}, concerned with maximizing predictive accuracy and employing sophisticated (but largely uninterpretable) ``black box'' machine learning models to this end. At the time, Breiman felt that the algorithmic culture was a troublingly small minority of statisticians -- a concern which may ring less true today when considering the rapid growth and adoption of machine learning which has at least partially penetrated the field of academic statistics. However, Breiman's distinction between these two cultures is still, to a large extent, visible in the respective techniques employed by each. This division between the \textit{data modeling culture} and the \textit{algorithmic culture} is clear to any reader of predictive modeling research. Both cultures contribute useful knowledge in the context of MOOCs: data models have the potential to inform course design and learning theory by revealing the underlying associations and mechanisms driving student outcomes; algorithmic models have the potential to support real-time early warning and intervention systems with highly accurate predictions even in the absence of interpretable knowledge about the underlying factors behind these predictions.

However, recent research in other fields, including the broader machine learning research community, has begun to erode the distinction between these two cultures, bringing us closer to having the best of both. Several streams of work have begun to make highly complex models more interpretable, gaining theory-building benefits without sacrificing the accuracy of those models. These include approximation approaches, which fit complex models and then approximate the final model using more interpretable linear \citep{Ribeiro2016-ut}) or decision tree models \citep{Craven1996-gq}, and perturbation approaches, which are used to inspect and explain individual predictions \cite{Baehrens2010-pa}. This work, along with others, suggests that predictive models are increasingly able to capture the benefits sought by the algorithmic culture -- notably, \textit{accurate} predictions of student success in MOOCs -- while also achieving the interpretability or theory-building results sought by the data modeling culture. Both Breiman and Domingos note that these more complex models typically fit the data \textit{better} (which is why they are preferred by the algorithmic culture) \cite{Breiman2001-py, Domingos1999-tl} --  and therefore an interpretable version of these models is likely to be \textit{more} informative and useful than the simple models traditionally used by data modelers, even for their own goals (understanding parameters and relationships in the data). Domingos argues that the notion that simpler models are preferable because simplicity is a goal in itself amounts to a mere \textit{preference} for simple models (which implies that the data modeling culture and the algorithmic culture simply have different preferences, but that neither approach is more ``correct'' \textit{a priori}). \cite{Domingos1998-dk, Domingos1999-tl} demonstrates that there is no trade-off between accurate and theory-building models: the notion that more interpretable models achieve better performance is demonstrably false under most conditions. 

Future research in predictive modeling in MOOCs should continue to explore techniques for making complex, highly accurate models more interpretable, following the lead of initial work by \cite{Nagrecha2017-dn}. This work is particularly salient in the case of educational student models, where the goal of such research is not only to understand the mechanisms underlying these models but also to \textit{intervene} to support students, and to actively support their achievement of certain outcomes (learning, sustained engagement, etc.). With a clear understanding of the patterns and relationships predictive models are identifying in MOOC data, many stakeholders in MOOCs would be able to act on this insights to support students. This includes course instructors, platform developers, course designers and content producers, support staff, community mentors, and even learners themselves. Furthermore, detailed inspection of models can help identify and reduce algorithmic bias in predictive student models \citep{Luo2015-mh}.

The dual advances in model interpretability and model fit do not, of course, absolve researchers from carefully considering the ethical implications of student models. To the contrary, as predictive models of student success improve and the use of their use becomes more widespread, the ethical implications of using these models -- and the responsibility of those constructing them -- will grow. Learning analytics and educational data mining researchers must consider and advocate for the use of student success models in ways that promote fairness, equity, and reductions in achievement gaps across student groups. This includes considering the training data itself and how (and whether) models based on this data might transfer to make predictions in other contexts or student populations, and working to prevent ``autopropaganda'' driven by such models \cite{Slade2013-zr}.

\subsection{Contributing to a Theory of Learning in MOOCs}~\label{sec:gaps-theory}

In terms of the three dimensions of predictive modeling research illustrated in Figure \ref{fig:modeling-dimensions} (accurate, actionable, and theory-building), the area where the research above has made the most limited contributions, relative to its potential, is in its contributions to theory, in particular to learning theory. 

We previously outlined how MOOCs represent a highly distinct domain for learners. While MOOCs require the development of novel learning theory for these novel contexts -- or at least the validation that traditional learning theory from brick-and-mortar environments, or similar digital learning environments such as e-learning, still hold in MOOC contexts. Predictive modeling research often utilizes an exceptional amount of learner data which is rarely available in more traditional educational environments. This data could be used not only to construct accurate or actionable models, which the field is making progress toward, but those which actually contribute to learning theory in novel ways. Indeed, a growing body of research has actively questioned whether the predictive component of predictive models is their most important contribution, instead arguing for more educational research which uses granular learning data to contribute to learning theory, not just make predictions \citep{Ho2017-qc}.

To date, this contribution has been limited, and while predictive modeling research may not always be able to support the types of rigorous causal inference necessary to serve as a foundation for learning theory, there is certainly more that such research can do to contribute to the development of learning theory. This includes grounding future predictive modeling efforts in known theoretical paradigms of student learning or engagement \citep[e.g.][]{Tinto2006-tj}.

\subsection{Understanding Long-Term Learner Success}

Connecting learners' course performance to anything outside the course is a challenge for future MOOC research to address. Little research has explicitly evaluated the connection between MOOC performance and future career or academic success (\cite{Wang2017-br} is a notable exception). Studies which evaluate real-world outcomes or link MOOC students to out-of-MOOC outcomes would be especially informative, because it is likely that such research more closely measures the outcomes we seek for many MOOC learners: it would be desirable for MOOC learners to experience career advancement and academic success outside of the platform, not for them to simply watch all course videos or persist until the end of a course (even though these are also useful, and relevant, outcomes in many cases). As we have previously mentioned, privacy protections to MOOC data and a reliance on optional learner questionnaires serve as barriers to this type of research. Using a diversity of available outcome measures (i.e. both engagement and learning) to evaluate existing predictive models would at least provide some indication of the all-encompassing learner success that these long-term outcomes represent, and is tractable with existing research methods and data.

\section{Conclusion}\label{sec:conclusion}

In less than a decade, MOOCs have emerged as a global source of educational opportunity and have reached millions of learners. Predictive models have been central to understanding user engagement and outcomes in MOOCs, and a diverse space of features and modeling techniques have been explored to this end. This includes diverse data sources, experimental subpopulations, feature extraction methods, modeling algorithms, prediction architectures, model evaluation techniques, and prediction outcomes. However, to date, little synthesis, survey, or critical evaluation of this work has been published. Such synthesis is necessary to survey the existing research and scientific consensus (or lack thereof) emerging in the field, and to direct future work. Furthermore, these student models can be used to actively support future MOOC learners, but only if they are sufficiently accurate, interpretable, and generalizable. This novel educational context requires a corresponding shift in research methodologies, which this survey demonstrates have achieved only incomplete adoption on the field of student success modeling to date. The MOOC research community stands to benefit substantially from the adoption of many of the techniques outlined in this paper.

In particular, our recommendations based on this work are that the field needs to move towards more robust model evaluation; broader experimental populations; and realistic experimental contexts. This will encourage growth toward building \textit{accurate}, \textit{actionable}, and \textit{theory-building} student success models. \textit{Actionable} models in particular are lacking in current MOOC research. We envision student success models supporting personalized, targeted interventions in MOOCs which are able to deliver effective support for students to reach a variety of goals. This vision can only be achieved, however, by closing the gaps outlined above: ineffective model evaluation will lead to poor generalizability and inaccurate identification of at-risk students; restrictive experimental sub-populations will yield models which are not applicable to large segments of learners; the use of unrealistic experimental contexts will produce models that are simply not operationalizable, requiring data which is not available at the time of prediction. However, if these barriers are overcome, MOOCs can truly deliver on their promise of providing effective educational opportunities for \textit{all} learners.

This is a critical time for the community to consider, and to repair, these methodological gaps. MOOCs, and the field of MOOC research, is transitioning from a nascent domain into a fully-fledged field of research, with canonical findings and scientific consensus beginning to emerge on key questions. However, failing to recognize where these gaps may affect scientific knowledge may result in this consensus forming around findings which have limited generalizability, methodological flaws, or practical barriers to implementation. Each of these gaps can be filled by adopting small changes to methodology in future research, and they can be further ameliorated by the construction of tools which enable researchers to follow these procedures without constructing the infrastructure themselves. In particular, we believe that future research which (a) replicates prior research using more rigorous statistical evaluation techniques, and (b) provides research tools and frameworks which support replication and benchmarking of published research, would be valuable contributions.

At the pace of current developments, we are optimistic about future developments in the field, and eager to see the impact these developments will bring to a generation of future MOOC learners.

\newgeometry{margin=1cm} 
\begin{landscape}
\fontsize{7}{9}\selectfont
\begin{longtable}{p{3cm}|p{2.75cm}p{1.5cm}p{0.5cm}p{0.5cm}p{2.5cm}p{6cm}p{2.5cm}p{2.5cm}}
\hline 
\textbf{Cite} & \textbf{Prediction \newline Outcome} & \textbf{Platform} & \multicolumn{2}{l}{\textbf{Courses/Students}} & \textbf{Data Source} & \textbf{Algorithms} & \textbf{Performance \newline Metrics} & \textbf{Prediction \newline Architecture} \\ \hline 
\endhead 
\cite{Kotsiantis2003-km} & Dropout & University &  & 354 & DEM; ASSGN & SVM; NN; NB;  LR; C4.5; KNN & ACC & Future course \\ \hline
\cite{Russo2005-nw} & Final grade & Other & 1 & 21 & FORUM & OLS & R2 & * \\ \hline
\cite{Wojciechowski2005-du}  & Final grade & Blackboard & 1 & 179 & SIS & CORR; OLS & Correlation; R2 & * \\ \hline
\cite{Cocea2007-eu} & Level of engagement & HTML Tutor; iHelp & 2 & 11 & CLICK & BN; LR; SL; IBk; ASC; BagCART; CVR; J48 & MAE; ACC; TPR; PREC & CV \\ \hline
\cite{Ramos2008-th} & Total exam points & WebCT & 2 & 119 & CLICK; ASSGN & OLS & R2 & T/T \\ \hline
\cite{Lykourentzou2009-ny} & Dropout & Moodle & 2 & 193 & DEM; ASSGN & PESFAM; NN; SVM; ENS & ACC; SENS; PREC & Future course \\ \hline
\cite{Garman2010-td} & Multiple grade-based metrics & Other & 1 & 235 & Cloze; ASSGN & LR & LogLik; K; GoF; CORR; DEV; HOSMER-LEMESHOW & * \\ \hline
\cite{Kotsiantis2010-hs} & Pass/Fail/Withdraw & Other & 1 & 1347 & SIS; LMS & KNN; NB; WINNOW; ENS; C4.5; NN; RIPPER; SVM; BP & ACC & \ldots \\ \hline
\cite{Barber2012-cc} & Failure & University of Phoenix & \ldots & \ldots & SIS; LMS & LR; NB; SVM; RF & None & T/T (model 1); CV (model 2) \\ \hline
\cite{Molina2012-xj} & Performance & Moodle & 32 & \ldots & SIS & MANY & SENS; PREC; F1; K; AUC & CV \\ \hline
\cite{Zafra2012-qa} & Final Grade & Moodle & 7 & 419 & CLICK; ASSGN & G3P-MI & SENS; SPEC; ACC & CV \\ \hline
\cite{Adamopoulos2013-pv} & Completion (Full/Partial/None) & Multiple & 133 & 842 & Reviews; META & RF; CART; SVM; LR; L2LR; & F1 & T/T \\ \hline
\cite{Balakrishnan2013-ty} & Dropout & edX & 1 & 29,882 & CLICK & HMM; ENS & AUC; LogLik; K; PREC; REC; F1; MCC & T/T \\ \hline
\cite{DeBoer2013-ki} & Course Grade & edX & 1 & 7,000 & DEM; ASSGN; CLICK & OLS & * & * \\ \hline
\cite{Ramesh2013-vu} & Certification & Coursera & 1 & 826 & CLICK; FORUM & PSL & AUC-PR (Pos, Neg); AUC; Kendall & CV \\ \hline
\cite{Romero2013-xe} & Final Grade & Moodle & 14 & 17,084 & SIS & MANY & ACC & CV \\ \hline
\cite{Yang2013-zq} & Dropout & Coursera & 1 & 771 & FORUM & SURV & None & * \\ \hline
\cite{Champaign2014-zz} & Assessment scores, skill, improvement & edX &  & 7140 & CLICK; Pre/Post Test; DEM & OLS; CORR & * & * \\ \hline
\cite{DeBoer2014-mq} & Assignment grades & edX & 1 & 30034 & CLICK & PANEL & None & * \\ \hline
\cite{Gutl2014-vx} & Dropout & \ldots & 1 & 1,680 & SURV & * & * & * \\ \hline
\cite{Halawa2014-pj} & Dropout & \ldots & 12 & \ldots & CLICK & THRESH & PREC; REC; SPEC & \ldots \\ \hline
\cite{Jiang2014-iv} & Achievement; certification & Coursera & 2 & 173,100 & FORUM & CORR & * & * \\ \hline
\cite{Jiang2014-vk} & Certification; Certification type & Coursera & 1 & 37,933 & ASSGN; FORUM; SIS & LR & ACC; AUC; PREC; REC; F1 & CV \\ \hline
\cite{Kloft2014-kb} & Dropout next week & \ldots & 1 & 20,828 & CLICK & SVM & ACC & CV \\ \hline
\cite{Ramesh2014-cy} & Performance; Persistence & Coursera & 1 & 826 & \ldots & PSL & AUC-PR Pos; AUC-PR Neg; AUC-ROC; Kendall & CV \\ \hline
\cite{Reich2014-ob} & Certification & HarvardX & 9 & 79,525 & DEM; SURV & LR; SURV & ACC; OR & * \\ \hline
\cite{Rose2014-mk} & Dropout & Coursera & 1 & 4,709 & FORUM & MMSB; SURV & None & * \\ \hline
\cite{Sharkey2014-tc} & Dropout & Coursera & 1 & 20000 & CLICK & RF & ACC; REC; TNR & T/T \\ \hline
\cite{Sinha2014-lw} & Engagement; Next click; In-video dropout; Course dropout & Coursera & 1 & 21952 & CLICK & L2LR; SURV & ACC; K; FNR & CV \\ \hline
\cite{Sinha2014-sr} & Dropout next week & Coursera & 1 & 14312 & CLICK; FORUM & NGRAM, SVM & ACC; K; FNR & CV \\ \hline
\cite{Stein2014-kx} & Dropout; video/quiz retention & Coursera & 1 & 35,819 & CLICK & COX & * & * \\ \hline
\cite{Taylor2014-hu} & Stopout & edX & 1 & 154,763 & CLICK; FORUM; ASSGN; WIKI; SURV; META; OTHER & LR, RLR & AUC; ACC & CV \\ \hline
\cite{Tucker2014-ue} & Quiz grades; assignment grades & Coursera & 1 & \ldots & FORUM; ASSGN & CORR & CORR & CORR \\ \hline
\cite{Veeramachaneni2014-ug} & Stopout & edX &  & 105,622 & CLICK; FORUM; ASSGN; WIKI; SURV; META; OTHER & LR & * & Predict ahead using "lag" \\ \hline
\cite{Wen2014-pv} & Dropout & Coursera & 3 & ,5507 & FORUM & SURV & * & * \\ \hline
\cite{Wen2014-xr} & Dropout from discussion forum & Coursera & 3 & 5,507 & FORUM & SURV & * & MULTIPLE \\ \hline
\cite{Yang2014-ka} & Dropout & Coursera & 3 & 5,507 &  & SURV & * & * \\ \hline
\cite{Ye2014-st} & Dropout week; Final grade & Coursera & 1 & 6,953 & \ldots & \ldots & ACC; SENS & \ldots \\ \hline
\cite{Ashenafi2015-fo} & Final Exam Grade & Other & 2 & 206 &  & OLS & RMSE & CV \\ \hline
\cite{Baker2015-uz} & Pass/Fail & Soomo & 1 & 4,002 & CLICK & W-J48; W-JRip; NB; W-KStar; SR; LR & PREC; REC; K; AUC & CV \\ \hline
\cite{Brinton2015-ya} &  Questions Correct on First Attempt (CFA) & Coursera & 1 & 5,205 & CLICK; ASSGN & MP; BIAS; MF; KNN; SVM & ACC; RMSE; AUC & CV \\ \hline
\cite{Brinton2015-bx} & Questions Correct on First Attempt (CFA) & Coursera & 2 & 6,450 & CLICK; ASSGN & DP; DT; CP; CT; SKR & ACC; F1 & CV \\ \hline
\cite{Brooks2015-ej} & Pass/Fail & Coursera & 4 &  & CLICK & J48 & ACC; K; REC; TNR; FNR & Future course \\ \hline
\cite{Boyer2015-lo} & Stopout & edX & 1 & 235,197 & CLICK; ASSGN & L2LR & AUC & MULTIPLE \\ \hline
\cite{Brooks2015-fh} & Completion & Coursera & 1 & 10,000 & CLICK; SURV & J48 & K & Future course \\ \hline
\cite{Chaplot2015-gp} & Dropout & Coursera & 1 & \ldots & CLICK; FORUM & NN & ACC, K, FNR & \ldots \\ \hline
\cite{Chaplot2015-dk} & Dropout & Coursera & 1 & \ldots & CLICK; FORUM & NN; HMM; RF; L2LR & K; FNR; ACC & CV \\ \hline
\cite{Coleman2015-cq} & Certification & edX & 1 & 43,758 & CLICK & LDA & Perplexity; ACC; REC; TNR & CV \\ \hline
\cite{Dowell2015-ip} & Final grade, social network centrality & edX & 1 & 1,754 & FORUM & MeM & AIC; LogLik; K; R2 & * \\ \hline
\cite{Fei2015-ea} & Dropout (multiple) & Coursera, edX & 2 & 39,877; 27,629 & CLICK & HMM; RNN; LSTM; SVM; LR & AUC & CV \\ \hline
\cite{Greene2015-no} & Total exam points; dropout current week & Coursera & 1 & 3875 &  & SURV; OLS & * & * \\ \hline
\cite{He2015-ab} & Failure & Coursera & 1 & 10,000 & CLICK; ASSGN & LR-SEQ; LR-SIM; SVM; RF; J48; NB; BN & AUC; SMOOTH & Future Course \\ \hline
\cite{Kennedy2015-jw} & Total course points & Coursera & 1 & 7,409 & CLICK; ASSGN & OLS & * & * \\ \hline
\cite{Kizilcec2015-gu} & Persistence; assignment grades & Coursera & 20 & 10,510 & CLICK & LR & AUC & CV \\ \hline
\cite{Koedinger2015-ef} & Dropout; final exam score & Coursera & 1 & 27720 & CLICK; DEM; OLI/ASSGN & LR; OLS & * & * \\ \hline
\cite{Sinha2015-yj} & Grade sequences & edX & 13 & 10,000 & CLICK; FORUM & CRF; LR; SMO & PREC; REC; F1 & T/T \\ \hline
\cite{Tang2015-ai} & Dropout & Harvardx-MITx & \ldots & 600,000 & CLICK; FORUM & DT & ACC & RESAMP \\ \hline
\cite{Wang2015-oy} & Learning gains & Coursera & 1 & 491 & CLICK; FORUM; ASSGN & OLS & * & * \\ \hline 
\cite{Whitehill2015-ap} & Stopout & HarvardX & 10 & \textgreater40,000 &  & L2LR & AUC & MULTIPLE \\ \hline
\cite{Yang2015-gy} & Dropout; confusion; confusion type & Coursera & 2 & 251 & CLICK; FORUM & SURV; LR & ACC; K & CV \\ \hline
\cite{Ye2015-uf} & Dropout & Coursera & 2 &  & CLICK & LR; SVM; CART; RF & F1 & \ldots \\ \hline
\cite{Ashenafi2016-df} & Course grade & Other & 2 & 229 & PEER & OLS & RMSE; FPR; TNR; ACC; ACC-W1; PREC; REC & Future course \\ \hline
\cite{Crossley2016-ij} & Completion & Coursera & 1 & 320 & CLICK; FORUM & DFA & ACC; F1; K & MULTIPLE \\ \hline
\cite{Dillon2016-fa} & Dropout & edX & 1 & 3,591 & CLICK; OTHER & * & PC & * \\ \hline
\cite{Evans2016-gj} & In-course engagement; persistence; completion & Coursera & 44 & 2.1M & CLICK; META & OLS & R2 & * \\ \hline
\cite{Joksimovic2016-mp} & Achievement; Certification & Coursera & 1 & 84786 & FORUM & ERGM; LR & AIC & * \\ \hline
\cite{Li2016-ze} & Dropout & XuetangX & 39 & \ldots & CLICK & LR; SVM; NB; CART; MT & PREC; REC; F1 & MULTIPLE \\ \hline
\cite{Li2016-gx} & Quiz scores & XuetangX & 1 & 2,633 & CLICK; ASSGN; DEM & MP; OLS; NN & MAE & \ldots \\ \hline
\cite{Liang2016-us} & Dropout & XuetangX & 39 & 20,000 & CLICK; META; ENRL & SVM; LR; RF; GBM & AUC & CV \\ \hline
\cite{Qiu2016-ct} & Certification; assignment grade & XuetangX & 11 & 88,112 & CLICK & LadFG; LR; SVM; FM & AUC; PREC; REC; F1 & \ldots \\ \hline
\cite{Ren2016-tr} & Homework/Quiz Scores & edX & 3 & 8,034 & CLICK & PLR, KT-IDEM & RMSE; ACC & Predict on train data \\ \hline
\cite{Robinson2016-yr} & Completion & HarvardX & 1 & 1,730 & SURV & L1LR & AUC & CV \\ \hline
\cite{Sanchez-Santillan2016-ml} & Pass/Fail & Moodle & 1 & 195 & LMS & JRIP; J48; C4.5 & ACC & CV \\ \hline
\cite{Vitiello2016-zn} & Dropout & Other & 5 & 2,056 & CLICK & K-MEANS; SVM & F1 & CV \\ \hline
\cite{Wang2016-yc} & Dropout & XuetangX & 39 & 79,186 & CLICK & LR; LR-SIM; LSTM; NSSM & AUC & T/T \\ \hline
\cite{Xing2016-le} & Dropout next week & Canvas & 1 & 3,617 & CLICK & GBN; C4.5; ENS & AUC; PREC & CV \\ \hline
\cite{Xu2016-pn} & Certification & HarvardX-MITx & 10 & \ldots & CLICK; FORUM & SVM & ACC & T/T \\ \hline
\cite{Bote-Lorenzo2017-yh} & Engagement $\pm$ & edX & 1 & 6,752 & \ldots & LR; SGD; RF; SVM & AUC & \ldots \\ \hline
\cite{Bouzayane2017-mw} & At-risk vs. active learners & \ldots & 1 & 1,535 & CLICK; FORUM & DRSA & PREC; REC; ACC & RESAMP \\ \hline 
\cite{Chen2017-sf} & Dropout & Coursera & 2 & 13,683 & CLICK & RF & F1 & \ldots \\ \hline
\cite{Garcia-Saiz2017-ed} & Pass/Fail & Moodle & 30 & \ldots & ASSGN & MANY & ACC & CV \\ \hline
\cite{Hosseini2017-kp} & Correct programming submission & \ldots & 4 & 798 & ASSIGN; DEM; OTHER & PFA & ACC & RESAMP \\ \hline 
\cite{Li2017-yd} & Final grade & Mengke & 4 & 32896 & CLICK & LR & F1 & CV \\ \hline
\cite{Nagrecha2017-dn} & Dropout & edX & 1 & 14,809 & CLICK & DT; RF; LR; GBT & AUC & CV \\ \hline
\cite{Vitiello2017-au} & Dropout & edX; Other & 13 & 38630 & CLICK & SVM; BDT & ACC & SSS \\ \hline
\cite{Vitiello2017-fr} & Completion; Completion vs. intent & Other & 11 & 3213 & CLICK & SVM & F1 & T/T \\ \hline
\cite{Whitehill2017-tt} & Dropout & HarvardX & 40 & 528,349 & CLICK; DEM; SURV & L2LR; NN & AUC & MULTIPLE \\ \hline

\caption{Literature review matrix of predictive modeling MOOC research. Ellipsis (\ldots) represents missing information that was not reported or not apparent upon review of published work. Asterisk (*) represents fields that are not applicable to a study. MANY indicates $\geq 10$ unique values for a given cell. Abbrevions are listed in Table \ref{tab:codes-table} below.}~\label{tab:master-matrix}
\end{longtable}
\end{landscape}
\restoregeometry

\begin{table}[]
\centering
\begin{tabular}{l p{8cm}} 
\hline 
\textbf{Field} & \textbf{Common Abbreviations} \\ \hline 
\textbf{Data Source} & CLICK = Clickstream; FORUM = Forum Posts; ASSGN = Assignments;  SIS = Student Information System; LMS = Learning Management System; DEM = Demographics; SURV = Survey; META = Course Metadata \\
\textbf{Algorithms} & LR = Logistic Regression; RF = Random Forest; OLS = Ordinary Least Squares Linear Regression; SURV = Survival Model; NN = Neural Network; NB = Naive Bayes; L2LR = L2-Penalized Logistic Rergession; ENS = Ensemble; BN = Bayesian Network \\
\textbf{Performance Metrics} & ACC = Accuracy; AUC = Area Under Receiver Operating Characteristic Curve; PREC = Precision; REC = Recall; K = Kappa; CORR = Correlation; DEV = Deviance \\
\textbf{Prediction Architecture} & T/T = Independent Train/Test Split; CV = Cross-Validation; CORR = Correlation Analysis; RESAMP = Resampling; SSS = Stratified Shuffle Split; * also used to code regression models where performance is evaluated directly on training data \\ \hline 
\end{tabular}
\caption{Abbreviations used in literature review matrix (Table \ref{tab:master-matrix}).}
\label{tab:codes-table}
\end{table}

\begin{acknowledgements}
This work was funded in part by the Michigan Institute for Data Science (MIDAS) Holistic Modeling of Education (HOME) project, and the University of Michigan Third Century Initiative. The authors would like to thank the four anonymous reviewers for their comments on the work.
\end{acknowledgements}

\bibliographystyle{spbasic}      
\bibliography{MSI_0_1}   

\bigskip 

\textbf{Josh Gardner} is a graduate student whose research centers on supporting and understanding data-driven learning at scale. His work includes statistical methods, applied research, and software development for large-scale statistical modeling in a variety of educational contexts, including MOOCs and residential higher education courses.

\bigskip 

\textbf{Christopher Brooks} is a Research Assistant Professor in the School of Information, and Director of Learning Analytics and Research at the Office of Academic Innovation at the University of Michigan. He is a Computer Scientist by background, and his work focuses on leveraging and supporting the diversity of students and their interactions in large scale online learning environments (e.g. MOOCs). His efforts include building models of educational discourse, predictive models of student success, and scaled replication infrastructure for educational data science. He teaches several large courses in the Applied Data Science with Python specialization on the Coursera platform.

\end{document}